\documentclass[
	journal
]{IEEEtran}

\IEEEoverridecommandlockouts % This is to make the \thanks{} command visible in conference mode
\makeatletter
\def\markboth#1#2{\def\leftmark{\@IEEEcompsoconly{\sffamily}\MakeUppercase{\protect#1}}%
\def\rightmark{\@IEEEcompsoconly{\sffamily}\MakeUppercase{\protect#2}}}
\makeatother

\renewcommand{\markboth}[1]{\renewcommand{\leftmark}{#1}\renewcommand{\rightmark}{#1}}
\usepackage[amssymb]{SIunits}
\usepackage[cmex10]{amsmath}
\usepackage{amssymb}
\usepackage{amsthm}
\usepackage[latin1]{inputenc}
\usepackage{cite}
\usepackage{url}
\usepackage{epstopdf}
\usepackage{glossaries}
\usepackage{graphicx}
\usepackage[small,bf]{caption}
\usepackage[subrefformat=parens,labelformat=parens]{subfig}
\usepackage{bm}
\usepackage[usenames]{color}
\usepackage[lined,ruled,linesnumbered]{algorithm2e}
\usepackage{balance}
\usepackage{array}
\usepackage{color}

\newcommand{\uA}{\mathsf{A}}
\newcommand{\setA}{\mathcal{A}}

\newcommand{\rxdeg}{r}

\newcommand{\contrib}{\psi}

\newcommand{\xqed}[1]{%
  \leavevmode\unskip\penalty9999 \hbox{}\nobreak\hfill
  \quad\hbox{#1}}
\newcommand{\exend}{\xqed{$\triangle$}}

\newenvironment{DIFnomarkup}{}{}
\newtheorem{example}{Example}

\newcolumntype{L}[1]{>{\raggedright\let\newline\\\arraybackslash\hspace{0pt}}m{#1}}
\newcolumntype{C}[1]{>{\centering\let\newline\\\arraybackslash\hspace{0pt}}m{#1}}
\newcolumntype{R}[1]{>{\raggedleft\let\newline\\\arraybackslash\hspace{0pt}}m{#1}}

\graphicspath{%
	{../pstricks/}%
}%

\newacronym{AWGN}{AWGN}{additive white Gaussian noise} 
\newacronym{CSA}{CSA}{coded slotted ALOHA} 
\newacronym{ABCSA}{B-CSA}{all-to-all broadcast CSA} 
\newacronym{SNR}{SNR}{signal-to-noise ratio} 
\newacronym{SINR}{SINR}{signal-to-interference-plus-noise ratio}
\newacronym{PLR}{PLR}{packet loss rate} 
\newacronym{UEP}{UEP}{unequal error protection} 
\newacronym{BS}{BS}{base station} 
\newacronym{LDPC}{LDPC}{low-density parity-check} 
\newacronym{VN}{VN}{variable node} 
\newacronym{CN}{CN}{check node} 
\newacronym{DE}{DE}{density evolution} 
\newacronym{MDS}{MDS}{maximum distance separable} 
\newacronym{BEC}{BEC}{binary erasure channel} 
\newacronym{PEC}{PEC}{packet erasure channel} 
\newacronym{DAMA}{DAMA}{demand assignment multiple access} 
\newacronym{CSMA}{CSMA}{carrier sense multiple access} 
\newacronym{VANET}{VANET}{vehicular ad hoc network} 
\newacronym{V2V}{V2V}{vehicle to vehicle} 
\newacronym{PHY}{PHY}{physical} 
\newacronym{MAC}{MAC}{medium access control} 
\newacronym{ARQ}{ARQ}{automatic repeat request} 
\newacronym{CDMA}{CDMA}{code division multiple access} 
\newacronym{TDMA}{TDMA}{time division multiple access}
\newacronym{CAM}{CAM}{cooperative awareness message}
\newacronym{DENM}{DENM}{decentralized environmental notification message}
\newacronym{ETSI}{ETSI}{European Telecommunications Standards Institute}
\newacronym{GPS}{GPS}{Global Positioning System}
\newacronym{DUEP}{DUEP}{double unequal error protection}
\newacronym{VC}{VC}{vehicular communication}
\newacronym{RU}{RU}{receiving user}
\newacronym{SIC}{SIC}{successive interference cancellation}
\newacronym{EF}{EF}{error floor}
\newacronym{ID}{ID}{induced distribution}

\newcommand{\eqsref}[2]{(\ref{#1})--(\ref{#2})}	
\newcommand{\figref}[1]{Fig.~\ref{#1}}

\newcommand{\tabref}[1]{Table~\ref{#1}}

\newcommand{\secref}[1]{Section~\ref{#1}}

\newtheorem{lemma}{Lemma}
\newtheorem{theorem}{Theorem}

\newtheorem{definition}{Definition}
\newtheorem{corollary}{Corollary}
\newtheorem{remark}{Remark}

\newcommand{\reva}[1]{{\color{black} #1}}
\newcommand{\revb}[1]{{\color{black} #1}}
\newcommand{\revc}[1]{{\color{black} #1}}
\newcommand{\revd}[1]{{\color{black} #1}}
\newcommand{\reve}[1]{{\color{black} #1}}
\definecolor{orange}{rgb}{1,0.5,0}
\newcommand{\revf}[1]{{\color{black} #1}}

\renewcommand{\max}[2]{\mathop{\mathrm{max}}_{#1} {#2}}
\renewcommand{\Pr}[1]{\mathrm{Pr}\left\{#1\right\}}
\newcommand{\expect}[2]{\mathsf{E}_{#1}\left\{#2\right\}}

\newcommand{\setU}{\mathcal{U}}

\newcommand{\setV}{\mathcal{V}}
\newcommand{\setC}{\mathcal{C}}
\newcommand{\setE}{\mathcal{E}}
\newcommand{\setS}{\mathcal{S}}
\newcommand{\setG}{\mathcal{G}}
\newcommand{\setGt}{\tilde{\mathcal{G}}}
\newcommand{\maxd}{q}
\newcommand{\lambdat}{\tilde{\lambda}}

\newcommand{\tframe}{t_{\text{frame}}}
\newcommand{\tslot}{t_{\text{slot}}}
\newcommand{\rdata}{r_{\text{data}}}
\newcommand{\packsize}{d_{\text{pack}}}
\newcommand{\tpream}{t_{\text{pream}}}
\newcommand{\tguard}{t_{\text{guard}}}
\newcommand{\tcsma}{t_{\text{csma}}}
\newcommand{\taifs}{t_{\text{aifs}}}
\newcommand{\tpack}{t_{\text{pack}}}

\begin{document}

\begin{DIFnomarkup}

\title{\revc{Broadcast Coded Slotted ALOHA: A Finite Frame Length Analysis}}

\author{
%Author 1, Author 2, Author 3, Author 4
Mikhail~Ivanov, Fredrik Br\"{a}nnstr\"{o}m, Alexandre Graell i Amat, and Petar Popovski
\thanks{This research was supported in part by the Swedish Research Council, under Grants No. 2011-5950 and 2011-5961, by the Ericsson's Research Foundation, by Chalmers Antenna Systems Excellence Center in the project `Antenna Systems for V2X Communication', and by the European Research Council, under Grant No. 258418 (COOPNET). The work of P. Popovski has been in part supported by the European
Research Council (ERC Consolidator Grant No. 648382 WILLOW) within the Horizon 2020 Program.
}
\thanks{M. Ivanov, F. Br\"{a}nnstr\"{o}m, and A. Graell i Amat are with the Department~of Signals and Systems, Chalmers University of Technology, SE-41296 Gothenburg, Sweden (e-mail: \{mikhail.ivanov, fredrik.brannstrom, alexandre.graell\}@chalmers.se).}
\thanks{Petar Popovski is with the Department of Electronic Systems, Aalborg University, 9220 Aalborg, Denmark (e-mail: petarp@es.aau.dk).}
\thanks{Parts of this work were presented at the IEEE International Conference on Communications (ICC), London, UK, June~2015.}
}

\maketitle

\end{DIFnomarkup}

%\begin{abstract}
%We propose an uncoordinated medium access control (MAC) protocol, called \emph{all-to-all broadcast coded slotted ALOHA (B-CSA)} for reliable message exchange between a large number of users in an all-to-all broadcast scenario. B-CSA is a very appealing protocol for transmitting periodic messages in vehicular networks. Unlike unicast coded slotted ALOHA (CSA), in B-CSA each user acts as both transmitter and receiver in a half-duplex mode. The half-duplex mode gives rise to an interesting  tradeoff expressed through a double unequal error protection (DUEP): the more the user repeats its packet, the higher the probability that this packet is decoded by other users, but the lower the probability for this user to decode packets from others. We analyze the performance of B-CSA over the packet erasure channel for a finite frame length. In particular, we provide a general analysis of stopping sets for B-CSA and derive an analytical approximation to the performance in the \gls{EF} region. The proposed analysis captures the DUEP feature of B-CSA. Simulation results reveal that the proposed approximation predicts very well the performance of B-CSA in the \gls{EF} region. Furthermore, it can be used to optimize the parameters of B-CSA. Finally, we compare the proposed B-CSA protocol with carrier sense multiple access (CSMA), currently adopted as the MAC protocol for vehicular networks. The results show that the proposed protocol is able to support a much lager number of users that can communicate reliably.
%\end{abstract}

\begin{abstract}
We propose an uncoordinated medium access control (MAC) protocol, called \emph{all-to-all broadcast coded slotted ALOHA (B-CSA)} for reliable all-to-all broadcast with strict latency constraints. In B-CSA, each user acts as both transmitter and receiver in a half-duplex mode. The half-duplex mode gives rise to a \emph{double unequal error protection (DUEP)} phenomenon: the more a user repeats its packet, the higher the probability that this packet is decoded by other users, but the lower the probability for this user to decode packets from others. We analyze the performance of B-CSA over the packet erasure channel for a finite frame length. In particular, we provide a general analysis of stopping sets for B-CSA and derive an analytical approximation of the performance in the \gls{EF} region, which captures the DUEP feature of B-CSA. Simulation results reveal that the proposed approximation predicts very well the performance of B-CSA in the \gls{EF} region. \revc{Finally, we consider the application of B-CSA to vehicular communications and compare its performance with that of carrier sense multiple access (CSMA), the current MAC protocol} in vehicular networks. The results show that B-CSA is able to support a much larger number of users than CSMA with the same reliability.
\end{abstract}

% Not needed for conference
%\begin{keywords} 
%	APSK constellation, nonlinear phase
%	noise, optical Kerr-effect, self-phase modulation. 
%\end{keywords}
% Redefine all acronyms that have been defined in the introduction
\glsresetall

\section{Introduction}\label{sec:intro}

\reve{Random access protocols based on slotted ALOHA~\cite{Abramson70, Roberts75} are widely used in wireless communication systems in order to support uncoordinated transmissions from a large number of users. These protocols offer low latency in scenarios in which each user is only intermittently transmitting.} \reva{ In slotted ALOHA, time is divided into slots and users select a single slot at random for transmission. If two packets are transmitted in the same slot, the respective receiver observes a \emph{collision} and the colliding packets are considered lost, which significantly limits the efficiency of slotted ALOHA.}

\reve{
In~\cite{Choudhury83}, it was suggested to repeat packets twice in randomly selected slots, thus slightly increasing the probability of a successful transmission. In~\cite{Casini07}, it was further suggested to utilize \gls{SIC}, as explained in the following. The system operates in \emph{frames}, where each frame is a periodically occurring structure that consists of a predefined number of slots. All users are assumed to be frame-synchronized. Each user transmits multiple copies (two or three) of its packet in a single frame, each copy in a different slot. Each copy of a packet contains pointers to all other copies of a packet. Once one copy is successfully received, the positions of the other copies are obtained and their interference in the respective slots is subtracted. Exploiting \gls{SIC} in~\cite{Casini07} provides significant performance improvement with respect to slotted ALOHA.} \revc{
\Gls{SIC} is also used in many other applications, e.g., \cite{Gollakota08, Tehrani11} to combat the hidden terminal problem in wireless networks, or in~\cite{ParandehGheibi10}, where it is combined with network coding.} 

\reve{
In~\cite{Liva11}, it was proposed to use different repetition factors for different users. To that end, users choose their repetition factor by drawing a random number according to a predefined distribution. It was recognized in~\cite{Liva11} that \gls{SIC} for the described protocol is similar to decoding of graph-based codes over the binary erasure channel. Hence, the theory of codes on graphs can be used to design good distributions.} \revc{ In~\cite{Narayanan12}, it was shown that using the so-called soliton distribution allows transmitting one packet in each slot when the frame length goes to infinity, which can be seen as the ``capacity'' of the protocol in~\cite{Liva11}. Coding over packets was used in~\cite{Paolini16} in the protocol termed \gls{CSA} in order to achieve high efficiency under transmit energy constraints.} \reve{ A protocol without a fixed frame structure was proposed in~\cite{Stefanovic13}.

The protocols in~\cite{Casini07, Gollakota08, Tehrani11, ParandehGheibi10, Paolini16, Stefanovic13} are designed for unicast transmission, i.e., when several users transmit to a common receiver (several receivers are possible in~\cite{ParandehGheibi10}).} \revc{In this paper, we consider a scenario where users exchange messages between each other, which is referred to as an \emph{all-to-all broadcast scenario}. This is a standard scenario used as a context for distributed consensus algorithms~\cite{Olfati-Saber07}. However, we chiefly draw our motivation from the emerging wireless scenario of \glspl{VC}, in which cars exchange safety messages. We propose a novel \gls{MAC} protocol based on \gls{CSA}, which we call \gls{ABCSA}.} \reve{ In particular, each user is equipped with a half-duplex transceiver, so that a user cannot receive packets in the slots it uses for transmission.\footnote{If full-duplex communication is possible, the analysis of the all-to-all broadcast scenario is identical to that of the unicast scenario, since each receiver operates undisturbed by its own transmissions.} The half-duplex mode gives rise to a \gls{DUEP} property: the more the user repeats its packet, the higher the chance for this packet to be decoded by other users, but the lower the number of available slots to receive in and, hence, the lower the chance to decode packets of others.

The proposed protocol provides a reliable access for a large number of devices under strict latency constraints, thus satisfying the needs of \glspl{VC} and other applications of future communications systems~\cite{Popovski14, Johansson15, Durisi16}. Since low latency is crucial in such applications, we analyze the performance of B-CSA in the finite frame length regime, which causes the appearance of an \gls{EF} in the performance of B-CSA.} \revc{ The \gls{EF} is due to stopping sets, which are harmful graph structures~\cite{Di02} that prevent iterative decoding.} \revd{Stopping sets are extensively analyzed for graph-based codes. In particular, stopping sets for a specific graph are well studied in~\cite{Wang09, Rosnes09} and references therein. CSA, on the other hand, is represented by a random graph and can be seen as a code ensemble. Stopping sets for code ensembles were analyzed in~\cite{Di02, Orlitsky05}, however, the obtained results are intractable for code lengths of interest and irregular codes. A similar approach to~\cite{Orlitsky05} was applied to~\gls{CSA} in~\cite{Pao15}, where quite loose bounds were obtained.} \reve{ In~\cite{Ivanov14} we proposed a low complexity \gls{EF} approximation for CSA at low-to-moderate channel load based on the heuristically determined ``dominant'' stopping sets, which is very accurate in the \gls{EF} region. Here, we extend the analysis in~\cite{Ivanov14} and propose a systematic way of determining dominant stopping sets together with their probabilities. The proposed analysis is able to capture the \gls{DUEP} feature of B-CSA. The  analytical approximation shows good agreement with the simulation results for low-to-moderate channel loads and can be used to optimize the parameters of B-CSA.

Finally, we compare the performance of B-CSA with that of \gls{CSMA}, currently adopted as the MAC protocol for \glspl{VC} over the \gls{PEC}. The use of the PEC is justified in that it provides a simplified model of the fading channel\cite{Fabregas06, Munari13, ParandehGheibi10}, which allows for a tractable analysis. Moreover, as we show in the paper, the all-to-all broadcast communication with half-duplex operation can be modeled as a \gls{PEC}. More accurate channel models were considered in~\cite{Herrero14}, however, they do not allow for a system optimization due to computational complexity. Our analysis shows that \gls{ABCSA} significantly outperforms \gls{CSMA} for channel loads of interest and that it is more robust to channel erasures due to the inherent time diversity.

The contributions are summarized in the following. (a) All-to-all broadcast \gls{CSA} is proposed;
(b) The analysis of CSA over the \gls{PEC} in~\cite{Ivanov14} is extended to the all-to-all broadcast scenario;
 (c) A more accurate analysis of the performance of \gls{CSA} compared to~\cite{Ivanov14} is presented, which includes a rigorous analysis of the probability of stopping sets and a systematic search of dominant stopping sets; 
(d) An analytical treatment of the \gls{DUEP} property for large frame lengths is presented; 
(e) A comparison of B-CSA with CSMA for \glspl{VC} over the PEC is carried out.
}

%For regular unicast \gls{CSA}, some simple approximations to characterize stopping sets were proposed in~\cite{Herrero14}.

%The analysis of \gls{CSA}-like schemes is usually done assuming that a packet in a collision-free slot can be reliably decoded and the interference caused by a packet can be ideally subtracted if the packet is known. Under these assumptions, the system can be viewed as a graph-based code operating over a \gls{BEC}~\cite{Liva11}. However, these assumptions do not capture some wireless channel effects such as fading. A simple way to deal with fading was suggested in~\cite{Stefanovic13}, but requires extra signalling from the \gls{BS} to estimate the channel. Such signalling is not possible in an all-to-all broadcast scenario. Other approaches as in~\cite{Herrero14} and~\cite{Stefanovic14} are much more involved and cannot rely on the \gls{BEC} model. 

\section{System Model}\label{sec:syst_model}

\revd{

We consider a network of $m+1$ users, which exchange messages between each other using half-duplex transceivers. We assume that time is divided into frames, users are frame synchronized,\footnote{The synchronization can be achieved by means of, e.g., \gls{GPS}, which provides an absolute time reference for all users.} and each user transmits one packet per frame. Frames are divided into $n$ slots, each slot matching the packet length.
}

\begin{figure}
	\centering
	\subfloat[Users' transmissions in a B-CSA system. Rectangles represent transmitted packets. The time slots in which user $\uA$ cannot receive are shown with gray. Erased packets due to the PEC are shown with hatched rectangles.]{\label{fig:system_model_a}
	\includegraphics{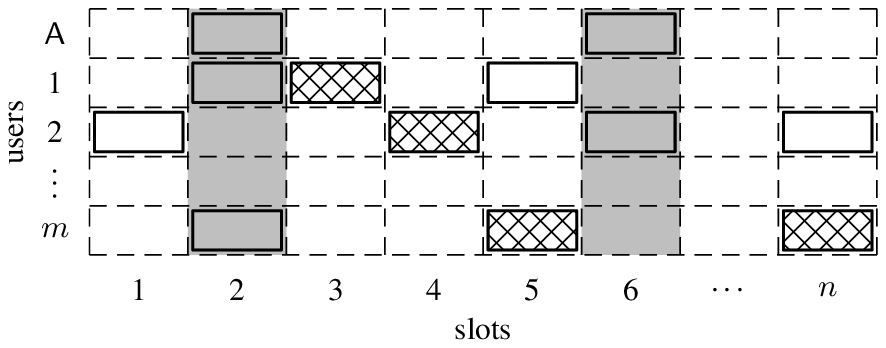}
	}	
	
	\subfloat[Original graph $\setGt$. The dashed lines correspond to the packets erased due to the PEC, i.e., the solid lines show the PEC induced graph $\setG$. ]{\label{fig:system_model_b}
	\includegraphics{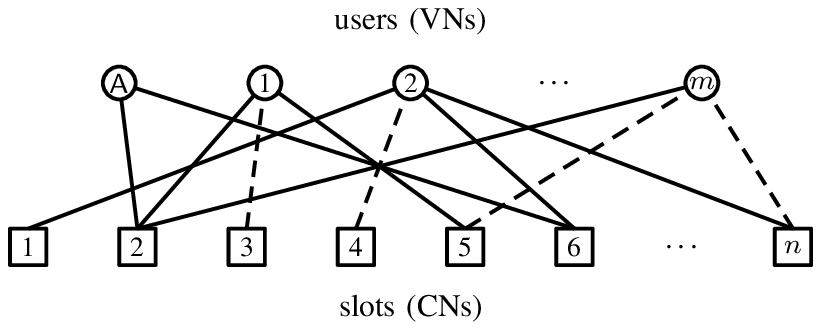}
	}
	
	\subfloat[PEC induced graph $\setG$. The dashed lines show the nodes and the packets erased due to broadcast, i.e., solid lines show the broadcast induced graph $\setG^{(\rxdeg)}$ for user $\uA$.]{\label{fig:system_model_c}
	\includegraphics{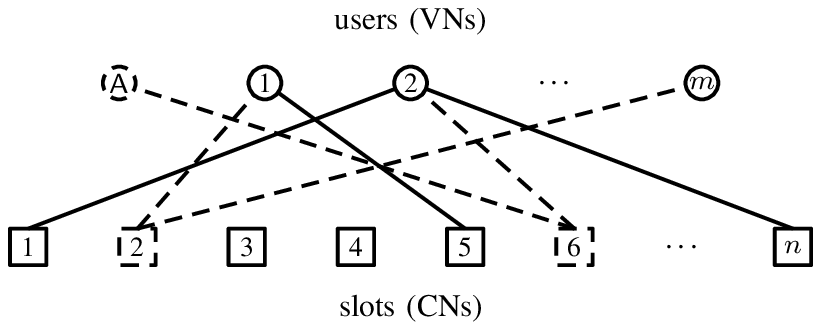}
	}
	\caption{System model.}	
	\label{fig:system_model}
\end{figure}

The transmission phase of the \gls{ABCSA} protocol is identical to that of unicast \gls{CSA}~\cite{Liva11}, and is briefly described in the following. Every user draws a random number $l$ based on a predefined probability distribution, maps its message to a PHY packet, and then repeats it $l$ times in randomly and uniformly selected slots within one frame, as shown in~\figref{fig:system_model}(a). Such a user is called a degree-$l$ user. Every packet contains pointers to its copies, so that, once a packet is successfully decoded, full information about the location of the copies is available.\footnote{\revd{The pointers can be efficiently represented by a seed for a random generator to reduce the overhead, as suggested in~\cite{Liva11}. We therefore ignore the overhead in this paper.}}

The main difference of the proposed \gls{ABCSA} protocol compared to unicast \gls{CSA} is that every user is also a receiver. Whenever a user does not transmit,  it buffers the received signal. Without loss of generality, we focus on the performance of a single user, denoted by $\uA$, also referred to as the receiver. $\mathcal{U}$ denotes the set of the other $m$ users, termed \emph{neighbors} of user $\uA$.

Since the frames are independent, it is sufficient to analyze the system within one frame. The received signal buffered by user $\uA$ in slot $i$ is 
\begin{equation*}
	y_i = \sum_{j \in \setU_i} h_{i,j} a_{j},
\end{equation*}
where $a_j$ is a packet of the $j$-th user in $\mathcal{U}$, $h_{i,j}$ is the channel coefficient \reva{between user $j$ and the receiver}, and $\setU_i \subset \mathcal{U}$ is the set of user $\uA$'s neighbors that transmit in the $i$-th slot. 

\revc{The $i$-th slot is called a \emph{singleton} slot if it contains only one packet. If it contains packets from more that one user, we say that a collision occurs in the $i$-th slot. Decoding proceeds as follows. First, user $\uA$ decodes the packets in singleton slots and obtains the location of their copies. We assume decoding is possible if the corresponding channel coefficient satisfies $|h_{i,j}| > C$, where $C$ is a threshold that depends on the physical layer implementation. If $|h_{i,j}| \le C$, we assume that the packet is erased, i.e., it cannot be decoded and does not cause any interference. The channel coefficients are assumed to be independent across users and slots and identically distributed such that $\Pr{|h_{i,j}| \le C} = \epsilon$ and $\Pr{|h_{i,j}| > C} = 1- \epsilon$, i.e., $\epsilon$ is the probability of a packet erasure. We refer to such a channel as a \gls{PEC}.}
\revd{Using data-aided methods~\cite{Lottici02}, the channel coefficients corresponding to the copies are then estimated.} \revc{ After subtracting the interference caused by the identified copies, decoding proceeds until no further singleton slots are found. We assume perfect interference cancellation, which is justified by physical layer simulation results in~\cite{Casini07, Liva11}. We remark that decoding is always performed in singleton slots, such that the code rates of different users do not have to satisfy the rate constraint for joint decoding~\cite{Rimoldi96}.
}

The system can be represented by a bipartite graph and can be analyzed using the theory of codes on graphs~\cite{Liva11}. In the graph, each user corresponds to a \gls{VN} and represents a repetition code, whereas slots  correspond to \glspl{CN} and can be seen as single parity-check codes. In the following, the terms ``users'' and ``\glspl{VN}'' are used interchangeably. A bipartite graph is defined by $\setG =({\setV}, {\setC}, {\setE})$, where $\setV$, $\setC$, and $\setE$ represent the sets of \glspl{VN}, \glspl{CN}, and edges connecting them, respectively. The number of edges connected to a node is called the node degree. An important parameter in the graph is the VN degree distribution \revc{~\cite{Luby98, Richardson01, Liva11}}
\begin{equation}\label{eq:distr_orig}
	\lambda(x) = \sum_{l = 0}^{\maxd}\lambda_{l}x^{l},
\end{equation}
where $\lambda_l$ is the probability of a VN to have a degree $l$ (i.e., the probability that the user transmits $l$ copies of its packet) and $\maxd$ is the maximum degree. We define a vector representation of~\eqref{eq:distr_orig} as $\bm{{\lambda}} = [{\lambda}_0, \dots, {\lambda}_q]$.\footnote{In this paper, we  consider only distributions with finite $q$ to avoid some technical problems in the following. This condition is always satisfied for all distributions of practical interest.} Furthermore, we define the graph profile as the vector $\bm{v}(\setG) = [ v_0(\setG), v_1(\setG), \dots, v_q(\setG)]$, where $v_l(\setG)$ is the number of degree-$l$ \glspl{VN} in $\setG$. The total number of VNs in $\setG$ is denoted by $\nu(\setG)$ and the total number of CNs is denoted by $\mu(\setG)$, \revd{i.e., $\nu(\setG) = m$ and $\mu(\setG) = n$.}

In this paper, we focus on the broadcast \gls{PLR}, \reva{defined as the probability that a user in $\setU$ is not resolved by the receiver. Since all users are independent, the PLR can be calculated as}
\begin{equation}\label{eq:plr_first}
\bar{p} = \frac{\bar{w}}{m},
\end{equation}
where $\bar{w}$ is  the average number of users that are not successfully decoded by user $\uA$, termed \emph{unresolved users}. Note that the PLR gives the probability that a packet of a user is not successfully received within a frame and does not refer to the copies sent by the user.
%\footnote{In fact, it would be more suitable to call this probability a message loss rate, but we have opted to use the more conventional term adopted in the random access literature.} 
 We define the channel load as the ratio of contending users and the number of slots, i.e., $g = (m+1)/n$. It should be noted that, in the unicast scenario, the channel load is calculated as $g = m/n$, since the receiver is not contending.

\section{Induced Distribution and Packet Loss Rate}\label{sec:analysis}

For transmission over a PEC, we showed in~\cite{Ivanov14} that the performance of unicast \gls{CSA} can be accurately \revd{approximated} based on an \emph{\gls{ID}} observed by the receiver. The fact that users in \gls{ABCSA}  cannot receive in the slots they use for transmission can also be modeled as packet erasures. Therefore, its performance can also be analyzed by means of the \gls{ID}. In this section, we derive the \gls{ID} in the general case of \gls{ABCSA} over the \gls{PEC}. To this end, we first find the degree distribution after the PEC and then the degree distribution perceived by user $\uA$. \reva{Throughout the paper, $l$ and $d$ denote the original and the induced degrees of a user in $\setU$, respectively, and $r$ denotes the degree of the receiver.}

\subsection{Induced Distribution}\label{sec:ind_distr}
For \gls{ABCSA} over the PEC, three different graphs can be defined. The first one is the \emph{original graph}, denoted by $\setGt$, that contains the edges \reve{$\tilde{\setE} = \{e_{i,j}: 1\le i \le n, \,\forall j \in \setU_i\}$}. We call its degree distribution the \emph{original distribution} (the one used by the users for transmission) and denote it by $\lambdat(x)$. The original graph corresponds to that of unicast CSA~\cite{Liva11} and its distribution is in the hands of the system designer.

 The \emph{PEC induced graph}, denoted by $\setG$, includes only the edges $e_{i,j} \in \tilde{\setE}$, for which $|h_{i,j}| > C$. In other words, $\setG$ is obtained from $\setGt$ by removing the edges corresponding to the erased packets. \revd{Since all elements of $\setG$ are contained in $\setGt$, we call $\setG$ a subgraph of $\setGt$ and write $\setG \subset \setGt$~\cite[Ch.~1.4]{Bondy76_Book}.} The VN degree distribution of the PEC induced graph is called the \emph{PEC~\gls{ID}} and is denoted by $\lambda(x)$. The graph $\setG$ is what a base station in unicast \gls{CSA} would observe after the PEC. However, only part of this graph is available to user~$\uA$ due to the half-duplex operation. Assuming that user~$\uA$ selects degree $\rxdeg$, we denote its available subgraph by $\setG^{(\rxdeg)}$ and call it a \emph{broadcast induced graph}.
$\setG^{(\rxdeg)}$ can be obtained from $\setG$ by removing the $\rxdeg$ CNs corresponding to the slots where user $\uA$ transmits and their adjacent edges. We call the degree distribution of this graph the \emph{broadcast \gls{ID}} and denote it by $\lambda^{(\rxdeg)}(x)$. The number of check nodes in the broadcast induced graph, $n^{(\rxdeg)} = \mu(\setG^{(\rxdeg)}) = n - \rxdeg$, is called the \emph{induced frame length}. For the example in~\figref{fig:system_model}(a), $\setGt$, $\setG$, and $\setG^{(\rxdeg)}$ are shown in~Figs.~\ref{fig:system_model}(b) and~\ref{fig:system_model}(c).

%An \gls{ID} is the distribution of user degrees as the receiving user sees it. For instance, in~\figref{fig:system_model}(a), user 1 has a  degree three and transmits in slots 2, 3, and 6.  The packet in slot 3 is erased by the PEC. Moreover, user $\uA$ also transmits in slot 2, hence, user $\uA$ perceives user 1 as a degree-1 user. The same user may be perceived differently by different users, i.e., if user 1 is in the neighborhood of user 2, user 2 sees it as a degree-2 user.

We now derive the PEC ID. Let a user from the set $\setU$ repeat its packet $l$ times. Each copy of this packet is erased with probability $\epsilon$. Hence, its degree in graph $\setG$ is $k \le l$  with probability $\binom{l}{k} \epsilon^{l-k} (1-\epsilon)^{k}$. Averaging over the original distribution $\tilde{\lambda}(x)$ leads to the PEC \gls{ID}
\begin{equation*}\label{eq:induced_distribution_pre}
	\lambda(x) = \sum_{l = 0}^{\maxd}\tilde{\lambda}_{l} \sum_{k = 0}^{l} \binom{l}{k} \epsilon^{l-k} (1-\epsilon)^{k}x^{k},
\end{equation*}
which can be written in the standard form~\eqref{eq:distr_orig}, where
\begin{equation}\label{eq:lambda}
	\lambda_{l} = \sum_{k = l}^{\maxd}\binom{k}{l} \epsilon^{k-l} (1-\epsilon)^{l}\tilde{\lambda}_{k}.
\end{equation}
Note that $\lambda_l$ is the fraction of users of degree $l$ after the PEC.
%For example, the distribution~\cite{Liva11}
%\begin{equation}\label{eq:distr_example}
%\lambda(x) = 0.25 x^2 + 0.6 x^3 + 0.15x^8,
%\end{equation}
%will be turned into the distribution
%\begin{align}
%\lambda'(x) =&  2.4\cdot 10^{-4}  + 1.6\cdot 10^{-2}x^1 + 2.9\cdot 10^{-1}x^2\nonumber\\
%		& + 5.5\cdot 10^{-1}x^3  +  7.5\cdot 10^{-6}x^4  + 2.0\cdot 10^{-4}x^5\nonumber\\
%		 & +  3.1\cdot 10^{-3}x^6  + 2.9\cdot 10^{-2}x^7 + 1.2\cdot 10^{-1}x^8
%\end{align}
%by the \gls{PEC} with $p = 0.03$.
 
Assuming that user $\uA$ selects degree $\rxdeg$, another user that has degree $l$ after the PEC is perceived by user $\uA$ as a degree-$d$ user if its non-erased transmissions take place in $l-d$ slots that are also selected by user $\uA$, which occurs with probability $\binom{n-\rxdeg}{d}\binom{\rxdeg}{l-d}/\binom{n}{l}$. Given the constraint $0\le l-d \le \rxdeg$, the broadcast \gls{ID} $\lambda^{(\rxdeg)}(x)$ observed by user $\uA$ can be written as 
\begin{equation*}
	\lambda^{(\rxdeg)}(x) = \sum_{d = 0}^{\maxd}\lambda^{(\rxdeg)}_{d}x^{d},
\end{equation*}
 where
\begin{equation}
	\lambda^{(\rxdeg)}_d = \sum_{l = d}^{\min\{q, \rxdeg+d\}}   \frac{\binom{n-\rxdeg}{d}\binom{\rxdeg}{l-d}}{\binom{n}{l}}\lambda_{l}\label{eq:t_induced}
%	& =\sum_{l = d}^{\min\{q, \rxdeg+d\}}   \frac{\binom{n-\rxdeg}{d}\binom{\rxdeg}{l-d}}{\binom{n}{l}}\sum_{k = l}^{\maxd}\binom{k}{l} \epsilon^{k-l} (1-\epsilon)^{l}\tilde{\lambda}_{k}\label{eq:induced_final}
\end{equation}
is the fraction of users of degree $d$ as observed by user $\uA$ if it selects degree $\rxdeg$. Clearly, the broadcast \gls{ID} depends on $n$, as opposed to the original and the PEC \glspl{ID}. The IDs in~\cite[eq.~(5)]{Ivanov14} and \cite[eq.~(3)]{Ivanov15} are a special case of~\eqref{eq:t_induced} when $\rxdeg=0$ and $\epsilon = 0$, respectively.

\begin{example}\label{ex:induced_distr}
For the original distribution
\begin{equation} \label{eq:distr_example}
\lambdat(x) = 0.5 x^2 + 0.5 x^4,
\end{equation}
the PEC \gls{ID} for $\epsilon = 0.01$ is
\begin{equation}\label{eq:distr_example_pec}
	\lambda(x) =  0.00005+ 0.0099x + 0.49x^2 + 0.019 x^3 + 0.48 x^4.
\end{equation}
The broadcast IDs depend on $n$. \revc{For $n = 100$, the broadcast \glspl{ID} are
\begin{align}
\lambda^{(2)}(x) &= 0.0004 + 0.03x + 0.47x^2 + 0.06 x^3 + 0.44 x^4,\label{eq:ex_dist1}\\
\lambda^{(4)}(x) &= 0.001 + 0.05x + 0.46x^2 + 0.09 x^3 + 0.41 x^4.\label{eq:ex_dist2}
\end{align}
for a receiver degree 2 and 4, respectively.}
\exend
\end{example}

The coefficient in front of $\lambda_l$ in~\eqref{eq:t_induced} can be written as
\begin{equation*}
\frac{(n-\rxdeg)! (n-l)!}{n! (n - \rxdeg - d)!}\frac{\rxdeg! l! }{d! (\rxdeg - l + d)! (l- d)!}\propto n^{d- l},
\end{equation*}
i.e., it tends to zero as $n \to \infty$ if $l > d$. Since $l \ge d$,  it can be shown that for any finite $\rxdeg$, $\lambda_d^{(\rxdeg)} = \lambda_d$ for all $\rxdeg$ and $d$ when $n \rightarrow \infty$, i.e., $\lambda^{(\rxdeg)}(x) \approx \lambda(x)$. This means that the effect of the broadcast nature of communications is negligible if the number of slots is large enough.
% In this case, density evolution can be used to predict the performance of \gls{ABCSA} in the waterfall region based on the original degree distribution $\lambda(x)$.
However, when the number of slots is small, which is the case in delay critical applications, the difference between the PEC ID and the broadcast ID is significant, especially if $\rxdeg$ is large.

%The main peculiarity of the \gls{ID}, not considered in the standard \gls{CSA} analysis, is that it contains all degrees up to $\maxd$, including zero and one. Such distributions are not considered when unicast \gls{CSA} is discussed and require special treatment. %This implies that the \gls{PLR} exhibits an \gls{EF}, which is lowerbounded by $\lambda_0^{(t)}>0$. Hence, the threshold, i.e., the channel load below which the \gls{PLR} is zero when $n \rightarrow \infty$, predicted by \gls{DE} is zero for any original distribution $\tilde{\lambda}(x)$. 

%Moreover, \gls{DE} devised in~\cite{Liva11} does not correspond to the decoding algorithm for distributions with $\lambda_1 \neq 0$. In the following, we focus on the \gls{ID} $\lambda(x)$ and the induced graph $\setG = \{\setV, \setC, \setE\}$.

\subsection{Packet Loss Rate}

Let $p_d^{(\rxdeg)}$ denote the probability that a user with degree $d$ in the broadcast induced graph $\setG^{(\rxdeg)}$ is not resolved by a receiver of degree $\rxdeg$. We refer to $p_d^{(\rxdeg)}$ as the degree-$d$ PLR. It can be calculated as
\begin{equation}\label{eq:plr_tot}
p_d^{(\rxdeg)} = \frac{\bar{w}^{(\rxdeg)}_d}{\bar{m}^{(\rxdeg)}_{d}} = \frac{\bar{w}^{(\rxdeg)}_d}{m \lambda^{(\rxdeg)}_{d}},
\end{equation}
where $\bar{m}^{(\rxdeg)}_{d}$ and $\bar{w}^{(\rxdeg)}_d$ are the average number of total and unresolved users of degree $d$ in $\setG^{(\rxdeg)}$, respectively. For $d = 0$, $p_0^{(\rxdeg)} = 1$ and $\bar{w}^{(\rxdeg)}_0 = m \lambda^{(\rxdeg)}_0$. For other degrees, we show how $\bar{w}^{(\rxdeg)}_d$ can be approximated based on the broadcast \gls{ID} in the next section. The probability that a degree-$\rxdeg$ receiver cannot resolve a user is called the average PLR and can be obtained by averaging~\eqref{eq:plr_tot} over the broadcast \gls{ID} as
\begin{equation}\label{eq:plr_rx}
p^{(\rxdeg)} = \sum_{d = 0}^{q}\lambda_{d}^{(\rxdeg)} p_{d}^{(\rxdeg)}. % = \sum_{l = 0}^{q}\lambdat_{l} \tilde{p}_{l}^{(\rxdeg)}.
\end{equation}

The probability that the original degree-$l$ user is not resolved by a degree-$\rxdeg$ receiver can be obtained as
\begin{align}
	\tilde{p}_{l}^{(\rxdeg)} 
	%&= \sum_{k = 0}^{l} \binom{l}{k}\epsilon^{l-k} (1 - %\epsilon)^{k}\check{p}_k^{(\rxdeg)}\\
 	&=	\sum_{k = 0}^{l} \binom{l}{k}\epsilon^{l-k} (1 - \epsilon)^{k} \hspace{-0.7cm} \sum_{d = \mathrm{max}\{k -\rxdeg, 0\}}^{k} \hspace{-0.3 cm} \frac{\binom{n-\rxdeg}{d}\binom{\rxdeg}{k-d}}{\binom{n}{k}} p_{d}^{(\rxdeg)},\label{eq:plr_tx_rx}
\end{align}
by reversing the operations in~\eqsref{eq:lambda}{eq:t_induced}. Finally, the broadcast PLR in~\eqref{eq:plr_first} is obtained as
\begin{equation}\label{eq:plt_rx_aver}
\bar{p} = \sum_{\rxdeg = 0}^{q}\lambdat_{\rxdeg} p^{(\rxdeg)}.
\end{equation}
%From~\eqref{eq:plt_rx_aver}, \eqref{eq:plr_rx}, and \eqref{eq:plr_tot} follows a simple lowerbound on the average PLR attained for $g = 0$ due to completely erased users, $p \ge \sum_{t = 0}^{q}\lambdat_{t} \lambda_0^{(t)}$.

From the equations above it is clear that the performance of a \gls{ABCSA} system depends on both the receiver and the transmitter degrees. We call this property \gls{DUEP} and formalize it in the following lemmas.

\begin{lemma}\label{theor:uep_tx}
For a given original distribution $\lambdat(x)$ and any $n$, $\tilde{p}_l^{(\hat{\rxdeg})} \ge \tilde{p}_l^{(\rxdeg)}$ if $\hat{\rxdeg} > \rxdeg$.
\end{lemma}
\begin{IEEEproof}
We prove the lemma by contradiction. Assume that the opposite holds, i.e., $\tilde{p}_l^{(\hat{\rxdeg})} < \tilde{p}_l^{(\rxdeg)}$ if $\hat{\rxdeg} > \rxdeg$. This implies that a degree-$\rxdeg$ receiver can improve its performance by ignoring $(\hat{\rxdeg}-\rxdeg)$ randomly selected slots, which cannot be true, as ignoring the slot information is the worst possible way to use the slot. This leads to a contradiction.
\end{IEEEproof}

Lemma~\ref{theor:uep_tx} describes the \gls{DUEP} from the receiver perspective. \revd{The \gls{DUEP} from the transmitter perspective is discussed in Lemma~2 in the following section.}

\section{Finite Frame Length Analysis}

From~\eqsref{eq:plr_tot}{eq:plt_rx_aver} it follows that the degree-$d$ PLR $p_d^{(\rxdeg)}$ is sufficient to describe all performance metrics for \gls{ABCSA}. $p_d^{(\rxdeg)}$ only depends on the distribution $\lambda^{(\rxdeg)}(x)$ and the induced frame length $n^{(\rxdeg)}$ seen by the receiver. The nature of these parameters is immaterial for the performance analysis. The receiver can be a user in a broadcast scenario that sees the broadcast ID $\lambda^{(\rxdeg)}(x)$. Alternatively, it can be thought of as a receiver in a unicast scenario in which the contending users use $\lambda^{(\rxdeg)}(x)$ as the original distribution with the frame length $n^{(\rxdeg)}$. Therefore, for the sake of simplicity, in this section we consider a unicast scenario, we omit superscript $(\rxdeg)$ and analyze $p_d$ in~\eqref{eq:plr_tot} for frame length $n$ and an arbitrary distribution $\lambda(x)$ used to generate graph $\setG$.

For $n \to \infty$, the typical performance of \gls{CSA} exhibits a threshold behavior, i.e., all users are successfully resolved if the channel load is below a certain threshold value, which can be  obtained via \gls{DE}~\cite{Liva11}. The threshold, denoted by $g^*(\bm{\lambda})$, depends only on the degree distribution $\lambda(x)$. 
\revd{We use the analysis in~\cite{Liva11} to describe the~\gls{DUEP} from the transmitter perspective in the following lemma.}

\begin{lemma}\label{theor:uep_rx}
\revd{For a given distribution $\lambda(x)$,} \revc{any load} \revd{$0 < g \le 1$, and sufficiently large $n$,  $p_{\hat{d}}< p_{d}$ if $\hat{d} > d$.}
\end{lemma}

\begin{IEEEproof}
\revd{We denote the degree-$d$ PLR at the $\rho$-th decoding iteration, $\rho \ge 1$, as a function of $n$ by $p_d(n, \rho)$.
According to the analysis in~\cite{Liva11, Lubi98},
\begin{equation} \label{eq:DE_UEP}
\lim_{n \to \infty} p_d(n, \rho) = (\xi_\rho)^d
\end{equation}
for any} \revc{finite} \revd{$d$, where $\xi_\rho$ is the probability of not removing an edge at the $\rho$-th decoding iteration. This probability is obtained recursively as~\cite[Sec.~III]{Liva11}
\begin{equation}\label{eq:DE_eq}
\xi_\rho = 1 -  \exp{\left(-g\lambda'(\xi_{\rho-1})\right)},
\end{equation}
where the prime denotes the derivative and $\xi_0 = 1$. \eqref{eq:DE_UEP} asserts that for any $\theta >0$ there exists an $n$ such that 
\begin{equation}\label{eq:de_ineq}
(\xi_\rho)^d - \theta \le p_d(n, \rho) \le (\xi_\rho)^d + \theta.
\end{equation}
It is easy to show from~\eqref{eq:DE_eq} that $\xi_\rho < 1$ for $0<g\le 1$ and any $\rho \ge 1$. Hence, $(\xi_\rho)^{\hat{d}}< (\xi_{\rho})^{d}$ and we can always find $\theta > 0$ such that $(\xi_\rho)^{\hat{d}} +\theta< (\xi_{\rho})^{d} - \theta$. Thus, according to~\eqref{eq:de_ineq} there exists an $n$ such that $p_{\hat{d}} < p_{d}$ at any iteration.}
\end{IEEEproof}
\revd{
\begin{remark}
In practice, it is sufficient that $n\gg q$ to guarantee $p_{\hat{d}} < p_{d}$ if $\hat{d}> d$.
\end{remark}}

\reva{
Lemma~\ref{theor:uep_rx} states that users that transmits more have a higher probability of being decoded by the receiver for sufficiently large frame lengths.
}
The finite frame length regime gives rise to an \gls{EF} in the PLR performance of \gls{CSA}. This \gls{EF} is due to stopping sets in the graph $\setG$. In this section, we first define stopping sets and analyze their contribution to the PLR. We then identify the  stopping sets that contribute the most to the \gls{EF} and propose an analytical  approximation to the performance in the \gls{EF} region.

\subsection{Stopping Sets and Their Contribution to Packet Loss Rate}
 
\begin{figure}
\centering
%\vspace{-0.34cm}
%	\centering
%	\subfloat[$\setS_{1}$.]{\label{fig:type1}
%	\makebox[1.5cm][c]{\includegraphics{../pstricks/letter/error_type11}}
%	}
	\subfloat[Stopping set $\setS$.]{\label{fig:two_two}
	\makebox[2cm][c]{\includegraphics{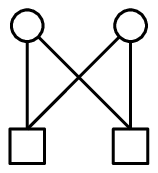}}
	}
	\subfloat[Graph $\setG$.]{\label{fig:graph_ex}
	\makebox[6.5cm][c]{\includegraphics{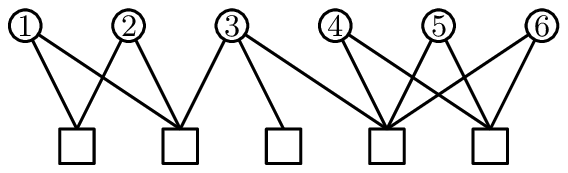}}
	}	
%	\subfloat[$\setS_{3}$.]{\label{fig:type2}
%	\includegraphics{../pstricks/letter/error_type112}
%	}	
%	\subfloat[$\setS_{4}$.]{\label{fig:type8}
%	\makebox[2cm][c]{\includegraphics{../pstricks/letter/error_type2}}
%	}
%	
%	\subfloat[$\setS_{5}$.]{\label{fig:type4}
%	\includegraphics{../pstricks/letter/error_type123}
%	}	
%	\subfloat[$\setS_{6}$.]{\label{fig:set222}
%	\includegraphics{../pstricks/letter/error_type3}
%	}	
%	\subfloat[$\setS_{7}$.]{
%	\includegraphics{../pstricks/letter/error_type223}
%	}
%	
%	\subfloat[$\setS_{8}$.]{
%	\includegraphics{../pstricks/letter/error_type1113}
%	}	
%	\subfloat[$\setS_{9}$.]{
%	\includegraphics{../pstricks/letter/error_type1122}
%	}
	\caption{\revd{Number of stopping sets $\setS$ in graph $G$.}}
%	\vspace{-0.4cm}
	\label{fig:error_events}
\end{figure}

Since erased packets are accounted for in the \gls{ID}, the only source of errors in the considered model is represented by the harmful structures in the graph $\mathcal{G}$. For example, when two degree-$2$ users transmit in the same slots (see \figref{fig:error_events}(a)), the receiver is not able to resolve them. Such harmful structures are commonly referred to as \emph{stopping sets}~\cite{Di02}. 
\begin{definition}
A connected bipartite graph $\setS$ is a stopping set if all \glspl{CN} in $\setS$ have a degree larger than one. 
\end{definition}

We say that a stopping set $\setS$ has profile $\bm{v}(\setS)$ and contains $\nu(\setS)$ VNs and $\mu(\setS)$ CNs. \revb{For example, for the stopping set in~\figref{fig:error_events}(a),  the graph profile is $\bm{v}(\setS) = [0, 0, 2,0,\dots, 0]$, where the number of zeros in the end depends on $q$, $\nu(\setS) = 2$, and $\mu(\setS) = 2$. Stopping sets are referred to as ``loops'' in~\cite{Herrero14}. However, stopping sets do not necessarily form a loop if degree-$1$ users are present. To analyze stopping sets, we define a VN-induced graph as follows.}

\revd{
\begin{definition}
A graph consisting of a subset of VNs of a graph $\setG$ and all their neighbors is called a VN-induced graph. 
\end{definition}
}

Since the graph $\setG$ and its profile $\bm{v}(\setG)$ are random, the average number of unresolved degree-$d$ users in~\eqref{eq:plr_tot} can be expressed using stopping sets as
\begin{equation}\label{eq:wl_super_general}
\bar{w}_d = \expect{\setG}{\sum_{\setS \in \setA}v_d(\setS) \hat{w}(\setS, \setG)},
\end{equation}
\revd{where $\setA$ is the set of all possible stopping sets for given $n$ and $m$, $v_d(\setS)$ is the number of degree-$d$ VNs in $\setS$, and $\expect{x}{\cdot}$ denotes the expectation over the random variable $x$. $\hat{w}(\setS, \setG)$ in~\eqref{eq:wl_super_general} is the number of VN-induced graphs in a graph realization $\setG$ which are both: i) isomorphic with $\setS$~\cite[Ch.~16.9]{Skiena08_Book}; ii) not a subgraph of a VN-induced graph isomorphic with another stopping set in $\setA$. With a slight abuse of notation, we refer to $\hat{w}(\setS, \setG)$ as the number of stopping sets $\setS$ in a graph $\setG$. The following example explains the definition of $\hat{w}(\setS, \setG)$.}

\revd{
\begin{example}
Consider the stopping set $\setS$ in~\figref{fig:error_events}(a) and the graph $\setG$ in~\figref{fig:error_events}(b). There are three VN-induced graphs in $\setG$ isomorphic with $\setS$, namely, the graphs induced by VNs $\{1,2\}$, $\{4,5\}$, and $\{5,6\}$. However, the two latter ones are subgraphs of another VN-induced graph isomorphic with a larger stopping set induced by the VNs $\{4,5,6\}$. Hence $\hat{w}(\setS, \setG) = 1$.
\end{example} 
}

The averaging in~\eqref{eq:wl_super_general} can be done in two steps as
\begin{equation}\label{eq:wl_general}
\bar{w}_d = \expect{\bm{v}(\setG)}{\sum_{\setS \in \setA} v_d(\setS) w(\setS, \setG)},
\end{equation}
where 
\begin{equation*}
	w(\setS, \setG) = \expect{\hat{\setG}: \bm{v}(\hat{\setG}) = \bm{v}(\setG)}{\hat{w}(\setS, \hat{\setG})}
\end{equation*}
is the average number of stopping sets $\setS$ in graphs with a particular profile $\bm{v}(\setG)$.

From the definition of a stopping set it follows that, for a given profile $\bm{v}(\setG)$, the number of stopping sets can be expressed as
\begin{equation}\label{eq:w4graph}
w(\setS, \setG) = \alpha(\setS, \setG) \beta(\setS) \gamma(\setS) \delta(\setS, \setG),
\end{equation}
where 
\begin{equation}\label{eq:multi_general}
\alpha(\setS, \setG) = \prod_{d = 1}^{\maxd} \binom{v_d(\setG)}{v_d(\setS)}
\end{equation}
is the number of ways to select VNs needed to create $\setS$ in a graph with profile $\bm{v}(\setG)$ and
\begin{equation}\label{eq:beta}
\beta(\setS) = \binom{n}{\mu(\setS)}
\end{equation}
is the number of ways to select $\mu(\setS)$ CNs out of $n$ CNs. $\gamma(\setS)$ in~\eqref{eq:w4graph} is  the probability of the selected VNs to be connected to the selected CNs so that $\setS$ is created. $\gamma(\setS)$ can be written as the ratio of the number of stopping sets $\setS$ that the selected VNs can create over the total number of graphs they can create, i.e.,
\begin{equation}\label{eq:gamma_general}
\gamma(\setS) = \frac{c(\setS)} {\prod_{d = 1}^{q}\binom{n}{d}^{v_d(\setS)}},
\end{equation}
where $c(\setS)$ is the number of graphs isomorphic with $\setS$ that the selected VNs can create. Unfortunately, deriving a closed-form expression for this constant does not seem to be straightforward. However, it can be found numerically according to its definition, as demonstrated in the following example.

\begin{example}\label{ex:gamma}
To find $c(\setS)$, we need to enumerate all combinations of connecting the VNs in $\setS$ to the CNs in $\setS$ so that the resulting graph is isomorphic with $\setS$. \revb{Consider stopping set $\setS$ in~\figref{fig:gamma_ex}(a). The only two graphs that are isomorphic with $\setS$ are shown in~Figs.~\ref{fig:gamma_ex}(b) and \ref{fig:gamma_ex}(c). Hence, $c(\setS) = 3$. To find $c(\setS)$, all graphs with a given profile $v(\setS)$ need to be generated and their isomorphism with $\setS$ tested~\cite[Ch.~16.9]{Skiena08_Book}.} \exend
\begin{figure}
\vspace{-0.34cm}
	\centering
%	\subfloat[$\setS_{\mathrm{I}}$ $c = 1$.]{\label{fig:type1}
%\includegraphics{../pstricks/letter/error_type3c}
%	}
	\subfloat[]{\label{fig:gamma_ex1}
	\includegraphics{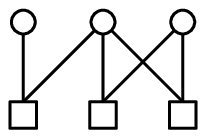}
	}
	\qquad
	\subfloat[]{\label{fig:gamma_ex2}
	\includegraphics{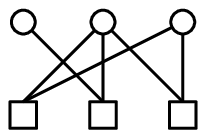}
	}
	\qquad
	\subfloat[]{\label{fig:gamma_ex3}
	\includegraphics{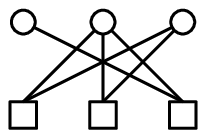}
	}
	\caption{\revb{Isomorphic graphs for a stopping set in~(a).}}
	\vspace{-0.4cm}
	\label{fig:gamma_ex}
\end{figure}
\end{example}

$\delta(\setS, \setG)$ in~\eqref{eq:w4graph} is the probability that the other $m-\nu(\setS)$ VNs are connected to CNs in such a way that another stopping set $\hat{\setS} \supset \setS$ is not created (see \figref{fig:error_events}). \revd{It does not have a closed-form expression in general and we resort to an upper bound.} \revd{By setting $\delta(\setS, \setG) = 1$ in~\eqref{eq:w4graph}, substituting the result into \eqref{eq:wl_general} and \revd{bringing} the expectation inside the summation, we obtain the upper bound given by
\begin{equation}
\bar{w}_d \le {\sum_{\setS \in \setA} v_d(\setS) \alpha(\setS) \beta(\setS) \gamma(\setS)},  \label{eq:ub}
\end{equation}
where $\alpha(\setS)= \expect{\bm{v}(\setG)}{\alpha(\setS, \setG)}$, which can be expressed as~\cite{Ivanov14}
\begin{equation}\label{eq:alpha_general}
	\alpha(\setS) =  \frac{m!}{(m - \nu(\setS))!} \prod_{d = 1}^{\maxd}\frac{\lambda_{d}^{v_{d}(\setS)}}{v_d(\setS)!}.
%\alpha(\setS) =  \binom{m}{\nu(\setS)}p_{\mathsf{mn}}(\bm{v}(\setS), \bm{\lambda}, \nu(\setS)).
\end{equation}
The upperbound in~\eqref{eq:ub} can be seen as a union bound, in which the occurrences of stopping sets are no longer exclusive events.

We now show that the bound in~\eqref{eq:ub} is tight for large $n$. $\delta(\setS, \setG)$ in~\eqref{eq:w4graph}} can be lower-bounded by the probability that none of these VNs is connected to the selected CNs. For a user of degree $d$, the probability of not being connected to the selected CNs is 
$\frac{\binom{n-\mu(\setS)}{ d}}{ \binom{n}{d}}$
for $n \ge d + \mu(\setS)$, which together with the expression for the binomial coefficient gives
\begin{equation}\label{eq:delta_tight_lb}
 \delta(\setS, \setG) \ge \prod_{d = 1}^{q} \left( \prod_{k = 0}^{d - 1} \frac{n - \mu(\setS) - k}{n - k} \right)^{(v_d(\setG) - v_d(\setS))^{+}},
\end{equation}
where $x^{+} = \max{}(0, x)$ and $n \ge q + \mu(\setS)$ (since the inequality $n \ge d + \mu(\setS)$ has to hold for any $d$). A looser lower bound can be obtained additionally assuming  that all these VNs have degree $q$,
\begin{align}
\delta(\setS, \setG) &\ge \left(\prod_{k = 0}^{q-1} \frac{n - \mu(\setS) - k}{n - k} \right)^{(m - \nu(\setS))^{+}}\nonumber\\
&\ge \prod_{k = 0}^{q-1} \left(  \frac{n - \mu(\setS) - q + 1}{n - q +1} \right)^{(m - \nu(\setS))^{+}}\nonumber\\
&= \left( 1 -  \frac{\mu(\setS)}{n - q +1} \right)^{q(m - \nu(\setS))^{+}} \nonumber\\ 
&\ge \left( 1 -  \frac{\mu(\setS)}{n - q +1} \right)^{qm} \nonumber\\ 
&\ge \exp{\left( -\frac{q \mu(\setS)m}{n - q + 1 - \mu(\setS)}\right)}, \label{eq:delta_lb}
\end{align}
where the last step follows from the inequality $\log(x) \ge 1- 1/x$ for $x > 0$. From~\eqref{eq:delta_lb} it follows that, for large $n$ ($n \gg q + \mu(\setS)$) and low channel loads ($n \gg q \mu(\setS) m$), $\delta(\setS, \setG) \approx 1$, \revd{which shows that~\eqref{eq:ub} is tight for large $n$.}

%%%%%%%%%%%%%%%%%%%%%%%%%%%%% Slotted ALOHA (example)
\begin{example}
For the distribution $\lambda(x) = x$ (slotted ALOHA), \revc{i.e., all VNs have degree 1,} all factors in~\eqref{eq:w4graph} can be easily calculated and the exact expression for~\eqref{eq:wl_general} can be obtained. Since all \glspl{VN} have degree 1, the profile of $\setG$ is deterministic with $\bm{v}(\setG) = [0,\,m]$, and the expectation in~\eqref{eq:wl_general} is trivial. Furthermore, all stopping sets have one CN with $s$,  $2 \le s \le m$, VNs connected to it. Such a stopping set, denoted by $\setS_s$, has parameters 
\begin{align*}
&\bm{v}(\setS_s) = [0,\,s],\quad \mu(\setS_s) = 1, \quad \alpha(\setS_s, \setG) = \binom{m}{s}, \nonumber\\ 
&\beta(\setS_s) = \binom{n}{1}, \, \gamma(\setS_s) = \phi^{s}, 
 \, \delta(\setS_s, \setG) = (1-\phi)^{m-s},
\end{align*}
where $\phi = 1/n$. We remark that the expression for $\delta(\setS_s, \setG)$ coincides with the bound in~\eqref{eq:delta_tight_lb} since the VNs ouside $\setS_s$ cannot be connected to $\setS_s$ without creating a larger stopping set and $c(\setS_s) = 1$ for any $s$.

%The CN degree distribution is  
%\begin{equation}\label{eq:aloha_cn}
%\rho_r = \binom{m}{r} \phi^{-r} (1 - \phi)^{m-r},
%\end{equation}
% $\rho_1$ is the probability that a \gls{CN} is connected to one VN and is equal to the throughput.\footnote{It is easy to see that in the asymptotic case, the throughput $\mathrm{th}$ is given by the well-known expression $\mathrm{th} = g \exp(-g)$.} All \glspl{CN} with degrees $r \ge 2$ correspond to stopping sets. The expected number of degree-$r$ \glspl{CN} (or, in other words, the expected number of the corresponding stopping sets) in the graph is equal to $n\rho_{r}$.
 
Since only degree-$1$ VNs are present in the graph, $\bar{w}_1$ (see~\eqref{eq:wl_general}) can be calculated exactly as
\begin{equation}\label{eq:w_aloha}
\bar{w}_1 = \sum_{s = 2}^{m} s \binom{m}{s} \binom{n}{1} 
\phi^{s} (1 - \phi)^{m-s}.
\end{equation}
Finally, by using the properties of the binomial distribution, we can obtain the exact expression for the \gls{PLR} in~\eqref{eq:plr_rx} as
\begin{multline}\label{eq:plr_sa}
p = p_1 = \frac{\bar{w}_1}{m} = \frac{n}{m} \sum_{s = 2}^{m} s \binom{m}{s} 
\phi^{s} (1 - \phi)^{m-s}\\
 = g^{-1} (m\phi - m\phi(1-\phi)^{m-1}) = 1 - (1-\phi)^{ng-1},
\end{multline}
which corresponds to the well known expression for the PLR of \revc{framed slotted ALOHA~\cite[eq.~(3)]{Lee05}}.\footnote{\revc{Eq.~(3) in~\cite{Lee05} gives the number successfully transmitting users.}}\exend

%We now examine the right-hand side of this expression. The summation over  $r = 2,\dots, m$ can be replaced by a summation over stopping sets that contain $r$ \glspl{VN} and one CN. Each such stopping set leads to $r$ unresolved users. $\binom{m}{r}$ is the number of combinations of $m$ VNs chosen $r$ times. $\binom{n}{1}$  is the number of combinations of $n$ slots chosen once. The next factor $\phi^{r}$ is the probability that the chosen $r$ VNs are connected to the chosen CN. The last factor $(1 - \phi)^{m-r}$ is the probability that all other $m-r$ VNs are not connected to the chosen CN. 
\end{example}

\subsection{Dominant Stopping Sets and Error Floor Approximation}\label{sec:dominant}

Identifying all stopping sets and calculating the corresponding $\gamma(\setS)$ in a systematic way for an arbitrary $n$ is not possible in general. We therefore determine stopping sets that contribute the most to the PLR for low channel loads, i.e., in the \gls{EF} region. It is clear that, for low channel loads ($n \gg m$), stopping sets with a small number of nodes are more likely to occur. We therefore focus on stopping sets with few CNs. Furthermore, to reduce the number of considered stopping sets, we define \emph{minimal} stopping sets as follows.
\begin{definition}
	A minimal stopping set is a stopping set that does not contain a nonempty stopping set of smaller size.  %Conversely, if some VNs can be removed so that the remaining graph is a stopping set, $\setS$ is said to be non-minimal. 
\end{definition}

\reva{It can be seen from~\eqref{eq:beta} and~\eqref{eq:gamma_general} that for low channel loads (more precisely, when $\sum_{d} d v_d(\setS) \ll n$), each edge in $\setS$ gives a factor $n^{-1}$ in~$\gamma$ and each CN gives a factor $n$ in $\beta(\setS)$. Assume now two stopping sets $\setS$ and $\hat{\setS}$ such that $\setS \subset \hat{\setS}$. To obtain $\setS$ from $\hat{\setS}$, more edges than CNs need to be removed from $\hat{\setS}$ since each CN in $\hat{\setS}$ is connected to at least two edges. Hence, the contribution of a non-minimal stopping set $\hat{\setS}$ to the PLR is smaller than that of $\setS$ for low channel loads. We therefore only consider minimal stopping sets in our analysis.} Simulation results in~\secref{sec:num_example} justify restricting to minimal stopping sets.

\reva{
We run an exhaustive search of minimal stopping sets with $\mu(\setS)$ up to five. For $\mu(\setS) \le 4$, there are 31 minimal stopping sets with the corresponding parameters given in~\tabref{tab:stopping_sets}. For $\mu(\setS) =  5$, there are 111 minimal stopping sets (not included in this paper). We remark that the complexity of finding $c(\setS)$ depends on the size of a stopping set~\cite[Ch.~16.9]{Skiena08_Book}. Therefore, we restrict the analysis to $\mu(\setS) \le 4$. The most dominant stopping sets presented in~\cite{Ivanov14} are a subset of the stopping sets in~\tabref{tab:stopping_sets}. Considering more stopping sets can improve the PLR approximation for moderate channel loads.
}

\begin{table}
\caption{Parameters of minimal stopping sets with $\mu(\setS)\le 4$. }
\begin{center}
\begin{minipage}{0.49\columnwidth}
  \begin{tabular}{C{1.7cm}| R{0.7cm} | R{0.6cm}}
	  \hline
$\bm{v}(\setS)$ & $\mu(\setS)$ & $c(\setS)$\\ 
    \hline
    \hline
$[0,\,2,\,0,\,0,\,0]$& $1$ & $1$\\
$[0,\,0,\,2,\,0,\,0]$& $2$ & $1$\\
$[0,\,2,\,1,\,0,\,0]$& $2$ & $2$\\
$[0,\,0,\,0,\,2,\,0]$& $3$ & $1$\\
$[0,\,1,\,1,\,1,\,0]$& $3$ & $3$\\
$[0,\,0,\,3,\,0,\,0]$& $3$ & $6$\\
$[0,\,0,\,2,\,1,\,0]$& $3$ & $6$\\
$[0,\,3,\,0,\,1,\,0]$& $3$ & $6$\\
$[0,\,2,\,2,\,0,\,0]$& $3$ & $12$\\
$[0,\,0,\,0,\,0,\,2]$& $4$ & $1$\\
$[0,\,1,\,0,\,1,\,1]$& $4$ & $4$\\
$[0,\,0,\,2,\,0,\,1]$& $4$ & $6$\\
$[0,\,0,\,1,\,2,\,0]$& $4$ & $12$\\
$[0,\,0,\,1,\,1,\,1]$& $4$ & $12$\\
$[0,\,0,\,0,\,3,\,0]$& $4$ & $24$\\
  \end{tabular}
  \end{minipage}
  \begin{minipage}{0.49\columnwidth}
  \begin{tabular}{C{1.7cm}| R{0.7cm} | R{0.6cm}}
$[0,\,0,\,0,\,2,\,1]$& $4$ & $12$\\
$[0,\,2,\,1,\,0,\,1]$& $4$ & $12$\\
$[0,\,2,\,0,\,2,\,0]$& $4$ & $24$\\
$[0,\,1,\,2,\,1,\,0]$& $4$ & $24$\\
$[0,\,1,\,2,\,1,\,0]$& $4$ & $24$\\
$[0,\,1,\,2,\,0,\,1]$& $4$ & $24$\\
$[0,\,1,\,1,\,2,\,0]$& $4$ & $48$\\
$[0,\,0,\,3,\,1,\,0]$& $4$ & $24$\\
$[0,\,0,\,3,\,0,\,1]$& $4$ & $24$\\
$[0,\,0,\,4,\,0,\,0]$& $4$ & $72$\\
$[0,\,0,\,3,\,1,\,0]$& $4$ & $144$\\
$[0,\,0,\,2,\,2,\,0]$& $4$ & $48$\\
$[0,\,0,\,2,\,2,\,0]$& $4$ & $48$\\
$[0,\,4,\,0,\,0,\,1]$& $4$ & $24$\\
$[0,\,3,\,1,\,1,\,0]$& $4$ & $72$\\
$[0,\,2,\,3,\,0,\,0]$& $4$ & $144$\\
	\hline	    
  \end{tabular}
  \end{minipage}
  \end{center}
  \label{tab:stopping_sets}
\end{table}

Constraining the set of the considered stopping sets in~\eqref{eq:ub} to minimal stopping sets with $\mu(\setS) \le 4$ and combining it with~\eqref{eq:plr_tot} gives the following approximation of the degree-$d$ PLR in the \gls{EF} region
\begin{equation}\label{eq:final_aprx}
	p_d \approx \frac{1}{m\lambda_d}\sum_{\setS \in \mathcal{A}_{31}}{v_{d}(\setS) \alpha(\setS) \beta(\setS)}\gamma(\setS),
\end{equation}
where $\mathcal{A}_{31}$ is the set of 31 minimal stopping sets in~\tabref{tab:stopping_sets} with $\alpha(\setS)$, $\beta(\setS)$, and $\gamma(\setS)$ given in~\eqref{eq:alpha_general}, \eqref{eq:beta}, and~\eqref{eq:gamma_general}, respectively. Since the considered stopping sets include VNs of degrees up to four, the approximation in~\eqref{eq:final_aprx} can only be used for $d \in \{1,\dots, 4\}$. If the PLR for larger degrees needs to be estimated, the set of considered stopping sets should be extended.

We remark that, in practice, distributions with large fractions of low-degree \glspl{VN} are most commonly considered since they achieve high thresholds. For instance, the soliton distribution~\cite{Narayanan12}, for which $g^*(\bm{\lambda}) = 1$, has $\lambda_2 = 0.5$ and  $\lambda_3 = 0.17$ (and $\lambda_0 = \lambda_1 = 0$). Moreover, IDs for \gls{ABCSA} and/or the \gls{PEC} have \revd{relatively high fractions of users of low degrees due to packet erasures}. We therefore conclude that the approximation in~\eqref{eq:final_aprx} is accurate for estimating the average performance of most \gls{CSA} systems of practical interest. This is supported by extensive numerical simulations, some of which are presented in the next section. %However, for distributions with large fractions of high degrees, e.g., $\lambda(x) = x^5$, the set of considered stopping sets may need to be extended.

\section{Numerical Results}\label{sec:num_example}
In this section, we first present different aspects of the \gls{ABCSA} performance for the distributions in Example~\ref{ex:induced_distr}. We then show how the proposed EF approximation can be used for the optimization of the original distribution.

\subsection{Induced Distribution and Packet Loss Rate}
\begin{figure}
\centering
\vspace{-0.35cm}
	\subfloat[$p_{d}^{(\rxdeg)}$.]{
		\includegraphics[]{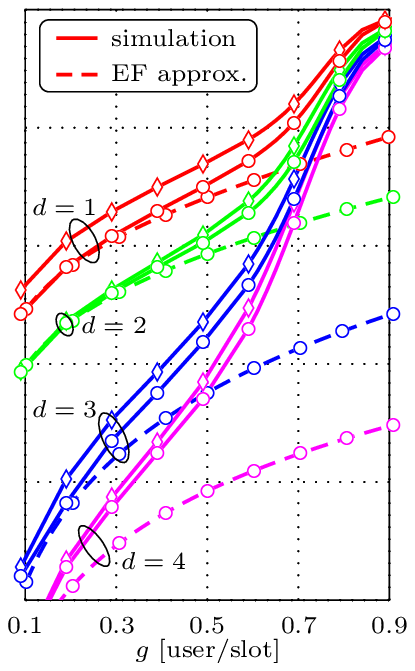}\label{fig:duepA}
	}
	\subfloat[$\tilde{p}_{l}^{(\rxdeg)}$.]{
		\hspace{-0.25cm}\includegraphics[]{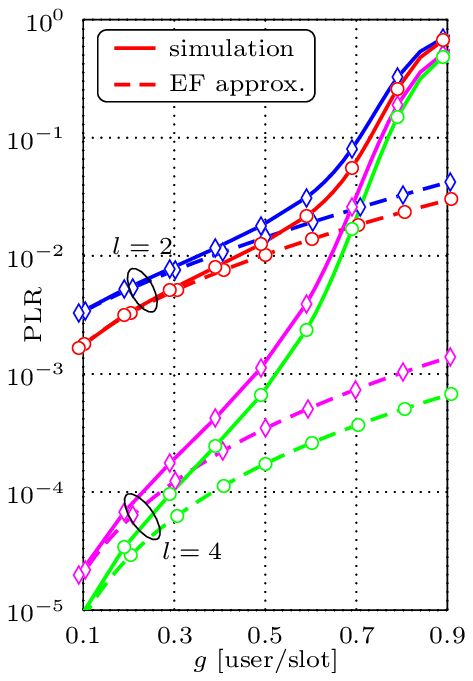}\label{fig:duepB}
	}
	\caption{PLR performance of B-CSA for the scenario in Example~\ref{ex:induced_distr} \revd{($n =100$)}. Solid curves show simulation results and dashed curves show analytical \gls{EF} approximations. Circles and diamonds show the PLR for $\rxdeg = 2$ and  $\rxdeg = 4$, respectively.}
	\label{fig:duep}
\end{figure}

In \figref{fig:duep}, we show the simulated PLRs $p_d^{(\rxdeg)}$ and $\tilde{p}_l^{(\rxdeg)}$ (solid lines) for the system described in Example~\ref{ex:induced_distr}. \figref{fig:duep}(a) shows the PLR for users of different induced degrees and illustrates the \gls{DUEP} property. Lines with circles show $p_d^{(2)}$ for the distribution in~\eqref{eq:ex_dist1} and lines with diamonds show $p_d^{(4)}$ for the distribution in~\eqref{eq:ex_dist2} and $d = 1,\dots,4$. It can be seen that for a given distribution, the larger the transmitter degree $d$, the better the performance, as  Lemma~\ref{theor:uep_rx} states.

\figref{fig:duep}(b) shows the simulated PLR $\tilde{p}_{l}^{(\rxdeg)}$ for the original transmitter degree $l$. As it can be seen from the figure, for a given $l$, the receiver with a smaller degree $\rxdeg$ has better performance in accordance with Lemma~\ref{theor:uep_tx}. On the other hand, for a given $\rxdeg$, the transmitter with a smaller degree $l$ has worse performance. The rationale behind the \gls{DUEP} is that the chance of a user to be resolved by other users increases if the user transmits more, but at the same time the chance to resolve other users decreases. 

We remark that the curves for $\tilde{p}_{l}^{(\rxdeg)}$ in \figref{fig:duep}(b) can be alternatively obtained from the solid curves in~\figref{fig:duep}(a) via~\eqref{eq:plr_tx_rx}. The resulting curves would then appear exactly on top of the simulation results in~\figref{fig:duep}(b), which confirms the correctness of~\eqref{eq:plr_tx_rx} and the derived IDs.

\subsection{Error Floor Approximation}

\figref{fig:duep} also shows the proposed analytical \gls{EF} approximation~\eqref{eq:final_aprx} (dashed lines). \figref{fig:duep}(a) shows~\eqref{eq:final_aprx} for  the distribution in~\eqref{eq:ex_dist1}, $\rxdeg = 2$ and $d = 1,2,3,4$, whereas \figref{fig:duep}(b) shows~\eqref{eq:plr_tx_rx} used together with the approximation~\eqref{eq:final_aprx} for $l = 2,4$ and $\rxdeg = 2,4$. The analytical \gls{EF} approximations demonstrate good agreement with the simulation results for low to moderate channel loads. This justifies the approximation in~\eqref{eq:final_aprx} and the use of minimal stopping sets. It can also be seen that the approximations for $d = 1,2$ are more accurate than those for $d = 3, 4$ since the stopping sets in \tabref{tab:stopping_sets} contain mostly users of low degrees. 

%\begin{figure}
%\centering
%\vspace{-0.35cm}
%	\subfloat[$p_{d}^{(2)}$.]{
%		\includegraphics[]{../pstricks/num_ex_a/broadcast_pec_a}
%	}
%	\subfloat[$p_{l}^{(t)}$.]{
%		\hspace{-0.25cm}\includegraphics[]{../pstricks/num_ex_b/broadcast_pec_b}
%	}
%	\caption{\gls{PLR} performance for Example~\ref{ex:induced_distr}. Dashed lines show simulation results and dash-dotted lines show analytical error floor approximations.}
%%	 \vspace{-0.4cm}
%	\label{fig:sim_results}
%\end{figure}

\begin{figure}
\centering
	\includegraphics[]{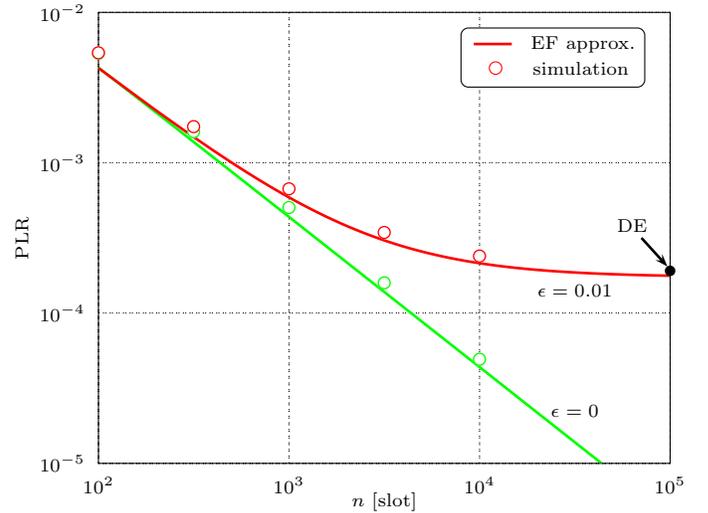}
	\caption{PLR performance of unicast CSA versus $n$ for the distributions in~\eqsref{eq:distr_example}{eq:distr_example_pec} and $g = 0.5$. Solid lines show the analytical approximation~\eqref{eq:final_aprx} with~\eqref{eq:plr_tx_rx} and markers show simulation results. The PLR value $2\cdot 10^{-4}$ predicted by DE is shown with a black dot.}
	\label{fig:plr_over_frame_length}
\end{figure}

\figref{fig:plr_over_frame_length} shows the dependency of the \gls{EF} on the frame length for the distributions in~\eqref{eq:distr_example} and \eqref{eq:distr_example_pec}, which correspond to $\epsilon = 0$ and $\epsilon = 0.01$, respectively,  in the unicast scenario.\revd{\footnote{\revd{For large values of $n$, we use the built-in approximation in the Matlab function \texttt{nchoosek} for the calculation of binomial coefficients.}}} It can be observed that without channel erasures (green curve), the \gls{EF} decays exponentially with $n$.  When erasures are present (red curve),  
the performance first decays exponentially for small $n$ and then approaches the value predicted by \gls{DE} (black dot) as $n$ grows. For  $n \lessapprox 10^3$, the finite frame length is the main cause of the \gls{EF}, whereas for $n \gtrapprox 10^3$ the PEC is the dominant factor causing the \gls{EF}. In this case, increasing $n$ does not improve the performance. Markers show simulation results, which agree well with the analytical approximation.

\begin{figure}
\centering
		\includegraphics[]{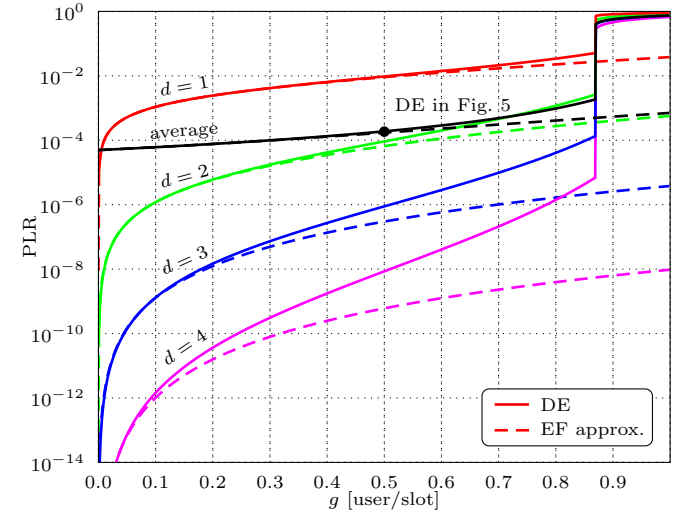}
	\caption{PLR performance of unicast CSA with the distribution in~\eqref{eq:distr_example_pec} and $n = 10^7$. Solid lines show the PLR predicted by DE and dashed lines show the proposed \gls{EF} approximation.}
	\label{fig:DE_UEP}
\end{figure}

Finally, in \figref{fig:DE_UEP} we show how the proposed \gls{EF} approximation compares with the DE results for large $n$. The solid lines show the PLR $\lim_{n \to \infty} p_d(n, \infty)$ obtained via DE~\eqref{eq:DE_UEP} for the distribution in~\eqref{eq:distr_example_pec} and $d = 1,\dots, 4$ in the unicast scenario. 
%In accordance with~Corollary~\ref{corol:DE}, the threshold for this distribution is zero. 
The proposed \gls{EF} approximation~\eqref{eq:final_aprx} for $n = 10^{7}$ is shown with dashed lines and agrees well with the DE curves. The lower the degree, the larger the range of channel loads for which the agreement is good. We remark that in all considered examples, the proposed analytical EF approximation always underestimates the PLR. This can be improved by including more stopping sets in~\eqref{eq:final_aprx}.

\subsection{Distribution Optimization for All-to-All Broadcast Coded Slotted ALOHA}

In this subsection, we concentrate on the broadcast scenario and discuss the optimization of the degree distribution for finite frame lengths using the proposed \gls{EF} approximation. To this end, we constrain the original distribution to have the form $\tilde{\lambda}(x) = \tilde{\lambda}_2 x^2 + \tilde{\lambda}_3 x^3 + \tilde{\lambda}_4 x^4 + \tilde{\lambda}_8 x^8$.  Such distributions have a good performance in the unicast scenario~\cite{Liva11} and are typical for low-density parity-check codes~\cite[p.~397]{Ryan_Lin_Book}.

Ideally, we would like to minimize the PLR around the values of $g$ at which the PLR curve switches from the  waterfall region to the \gls{EF} region. However, analytical tools to predict the PLR at such channel loads are missing. The proposed EF approximation~\eqref{eq:final_aprx} is accurate only for low to moderate channel loads. If the \gls{EF} approximation is the only optimization objective, the optimal distribution is always $\tilde{\lambda}(x) = x^8$. However, such a distribution has $g^*(\bm{\lambda}) = 0.54$, hence, bad performance at channel loads of interest. Here, we use a linear combination of the broadcast PLR in~\eqref{eq:plt_rx_aver} and the threshold as the optimization objective, which corresponds to scalarization of a multidimensional objective function~\cite[Ch.~4.7]{Boyd09_Book}.

\begin{figure}
\centering
	\subfloat[\gls{EF} vs. threshold tradeoff.]{
		\includegraphics[]{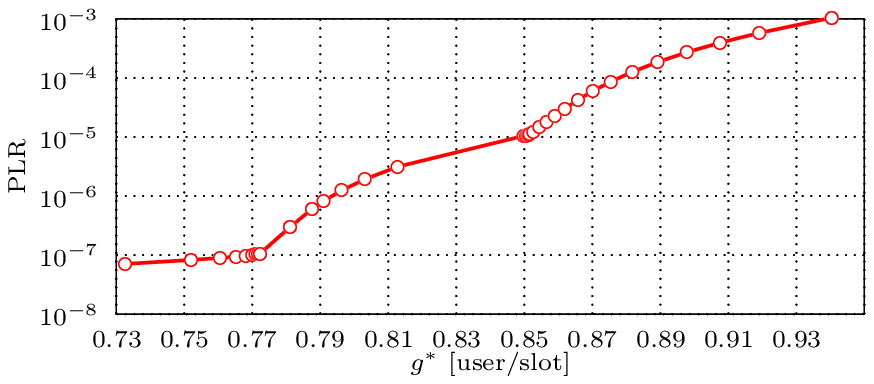}\label{fig:optim_a}
	}
	
	\subfloat[Optimal distributions.]{
		\includegraphics[]{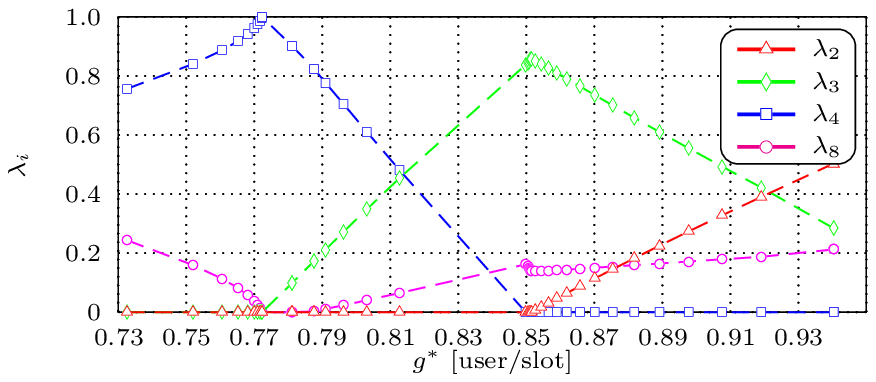}\label{fig:optim_b}
	}
	\caption{Distribution optimization for $n = 500$, $g = 0.5$, and $\epsilon = 0$.}
	\label{fig:tradeoff}
\end{figure}

For notational convenience, we write the broadcast PLR in~\eqref{eq:plt_rx_aver} as $\bar{p} (\bm{\tilde{\lambda}})$ to highlight that it depends on the distribution $\bm{\tilde{\lambda}}$ and formulate the following optimization problem
\begin{equation*}
\begin{aligned}
& \underset{\bm{\tilde{\lambda}}}{\text{minimize}}
& &  -g^*(\bm{\tilde{\lambda}}) + \eta \bar{p} (\bm{\tilde{\lambda}})\\
& \text{subject to}& & \tilde{\lambda}_i \ge 0  \; i = 2, 3, 4, 8,\\
& & & \tilde{\lambda}_i = 0  \; \text{otherwise},\\
& & & \lambdat_2 + \lambdat_3 + \lambdat_4 + \lambdat_8 = 1,\\
\end{aligned}
\end{equation*}
where $\eta$ is a weighting coefficient. We numerically solve this optimization problem\footnote{We  solve the optimization problem by means of the Nelder-Mead simplex algorithm~\cite{Lagarias98}. Global optimality is therefore not guaranteed.} by using the EF approximation~\eqref{eq:final_aprx} to calculate $\bar{p} (\bm{\tilde{\lambda}})$ for different values of $\eta$ and obtain the  \gls{EF} vs. threshold tradeoff shown  in~\figref{fig:tradeoff}(a). The corresponding optimal distributions are shown in~\figref{fig:tradeoff}(b). As it can be seen from~\figref{fig:tradeoff}(a), the optimal distributions around $g^* = 0.85$ provide relatively high threshold for relatively low \gls{EF} values ($\approx 10^{-5}$). We pick the distribution 
\begin{align}\label{eq:optim_distr}
\tilde{\lambda}(x) &=  0.86 x^3 + 0.14 x^8
\end{align}
and use it in the next section when we compare \gls{ABCSA} with \gls{CSMA}. \revf{We remark that the choice of $g^*$ for selecting the distribution depends on the required reliability. Since global optimality of the results in~\figref{fig:tradeoff} is not guaranteed, better distributions can potentially be obtained. However, the presented distributions are the best known distributions that provide the EF vs. threshold tradeoff.}

The results in~\figref{fig:tradeoff} are obtained for $n = 500$, $g = 0.5$, and $\epsilon = 0$. We remark that the distribution in \eqref{eq:optim_distr} is close to the one presented in~\cite{Ivanov14} for the unicast scenario. Furthermore, we observed that the optimization problem is not very sensitive to the choice of parameters $g$, $n$, and $\epsilon$.

\revc{

\section{Broadcast Coded Slotted ALOHA in Vehicular Networks} \label{sec:csma}

In this section, we evaluate the performance of B-CSA in a vehicular network and compare it with the currently used \gls{CSMA} protocol. The physical layer parameters used here are taken from~\cite{IEEE80211} and are given in~\tabref{tab:params}. We consider transmission of \gls{CAM}~\cite{etsi_cam} packets that are sent periodically every $\tframe$ seconds. We set the frame duration equal to $\tframe$ and we assume that the network does not change during this period. We assume that all packets have length $\tpack$, which depends on the packet size $\packsize$, transmission rate $\rdata$, and the length of the preamble added to every packet $\tpream$, i.e., $\tpack = \tpream + \packsize/\rdata$.} \revd{In addition, the slot duration $\tslot = \tpack + \tguard$, where $\tguard$ accounts for timing inaccuracy. The number of slots is determined as $n = \lfloor \tframe/ \tslot \rfloor$. The inclusion of a guard interval, $\tguard$, reduces the number of slots in a frame and thus worsens the performance of B-CSA.} 

\revc{
\begin{table}
\caption{The PHY parameters. The values in the upper part are taken from~\cite{IEEE80211}; the values in the lower part are derived. }
\begin{center}
  \begin{tabular}{l |c| c | c|c}
	  \hline
    Parameter& Variable & \multicolumn{2}{c|}{Value} & Units\\ 
    \hline
    \hline
    Data rate &$\rdata$& \multicolumn{2}{c|}{6}& Mbps\\
	\hline
	PHY preamble &$\tpream$& \multicolumn{2}{c|}{40}& $\mu$s\\
	\hline
	CSMA slot duration &$\tcsma$& \multicolumn{2}{c|}{13}& $\mu$s \\
	\hline
	AIFS time &$\taifs$ & \multicolumn{2}{c|}{58}& $\mu$s \\
	\hline
	\hline
	Frame duration& $\tframe$ & \multicolumn{2}{c|}{100}& ms\\
	\hline
	Guard interval &$\tguard$ & \multicolumn{2}{c|}{5}& $\mu$s\\
	\hline
	Packet size &$\packsize$& 200 & 400& byte\\
	\hline
	Packet length &$\tpack$& 312 & 576& $\mu$s\\
	\hline
	Slot duration &$\tslot$ & 317 & 581& $\mu$s\\
	\hline 
	Number of slots &$n$ & 315 & 172&\\
	\hline	    
  \end{tabular}
  \end{center}
  \label{tab:params}
\end{table}

\subsection{Carrier Sense Multiple Access}

\gls{CSMA} is used as MAC protocol for vehicular networks~\cite{IEEE80211}. We analyze the following system model that can be compared with B-CSA. We consider a network with $m+1$ users indexed by $j = 1,\dots, m+1$, where every user is within all other users transmission range, i.e., no collision occurs due to the hidden terminal problem.  $m$ can be thought of as the instantaneous number of neighbors for a given user. We assume that a user can always sense other users transmissions. Collisions are assumed to be destructive. If no collision occurs, each user may not able to decode a packet with probability $\epsilon$ due to noise-induced errors. We say that such a packet is erased. Note that the backoff protocol is not affected by channel erasures and partial collisions are not possible.
 
}
 
The set of users is denoted by $\mathcal{V}$ and time is denoted by $t$. At the beginning of the contention ($t = 0$), every user selects a real random number $\tau_j \in [0,\,\,\tframe)$, which represents the time when the $j$-th user attempts to transmit its first packet invoking the \gls{CSMA} procedure from~\cite[Fig.~2(a)]{Bilstrup09}. The contention window size is selected to be 511~\cite{Ivanov15}. \reva{$\taifs$ is the sensing period during which the users sense the channel to determine whether it is busy of not, where AIFS stands for arbitration interframe space}. $\tcsma$ is the duration of a backoff slot. The values of these parameters are specified in~\cite{IEEE80211} (see~\tabref{tab:params} for the values used in  simulations).

To overcome the effect of packet erasures, each user attempts $\kappa$ transmissions of each packet. At time instant $\tau_j + (k-1) \tframe/\kappa +  (i-1) \tframe$, the $j$-th user makes the $k$-th attempt, $k = 1,\dots, \kappa$, to transmit its $i$-th packet, $i = 1,\,2,\,\dots$.  If by the time $\tau_j + i \tframe$ none of the copies of the $i$-th packet is transmitted, the packet is dropped. For $\kappa = 1$, the described protocol is considered in~\cite{Ivanov15}.

The channel load is defined as the ratio of the number of users, $m+1$, and $\tframe$ (expressed in slots) to match the definition of the channel load for \gls{ABCSA}, i.e., $g = (m + 1)/(\tframe/\tslot) = (m+1)/n$. The \gls{PLR} for user $j$ is defined as 
\begin{equation*}
	p_j = 1 -  \frac{\expect{\tau_1,\dots, \tau_m}{\sum_{\substack{i \in \mathcal{V}\\ i\neq j}} \eta_{i,j}}}{ m},\label{eq:plr_csma}
\end{equation*}
where $\eta_{i,j} \in \{0, 1, 2 \}$ is the number of packets of user $i$ successfully received by user $j$ over the time interval $[t_0,\,\, t_0 + \tframe)$. To estimate the performance, we introduce a time offset $t_0 = 2\tframe$ in order to remove the transient in the beginning of the contention. As the performance does not depend on the particular user, without loss of generality, we select user $j = 1$.

\begin{figure}
\centering
		\includegraphics[]{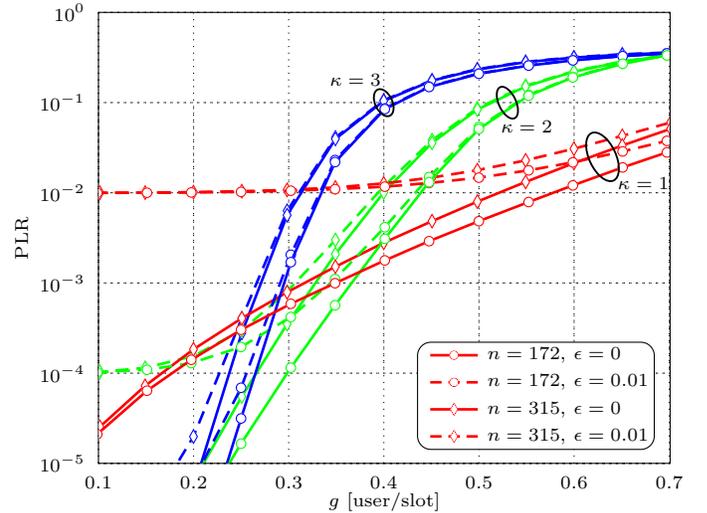}
	\caption{PLR performance for CSMA over the PEC. Solid  and dashed lines correspond to $\epsilon = 0$ and $\epsilon = 0.01$, respectively. Circles and diamonds show the PLR for $n = 172$ and $n = 315$, respectively.}
	\label{fig:csma_pec}
\end{figure}

The performance of CSMA  for different values of $\epsilon$, $\kappa$, and the frame lengths in~\tabref{tab:params} is shown in~\figref{fig:csma_pec}. We can observe that the performance of CSMA degrades as $n$ increases due to the increase of sensing overhead. Furthermore, for a given $n$, increasing  $\kappa$ reduces the achievable throughput but improves the performance at low channel loads. The PLR curves approach the value $\epsilon^{\kappa}$ for $g = 0$, hence, the number of repetitions can be predicted based on $\epsilon$ and the required reliability.

\subsection{Carrier Sense Multiple Access vs. All-to-all Broadcast Coded Slotted ALOHA}

Even though the reliability requirements are not specified in~\cite{IEEE80211} and depend on the particular application, for the sake of comparison we assume that the broadcast PLR of interest is in the range $10^{-2}\, -\, 10^{-3}$. From~\figref{fig:csma_pec} it follows that for CSMA, $\kappa = 2$ provides good performance for the PEC with $\epsilon \le 10^{-2}$. For \gls{ABCSA} we select the distribution in~\eqref{eq:optim_distr}.

\begin{figure}
\centering
		\includegraphics[]{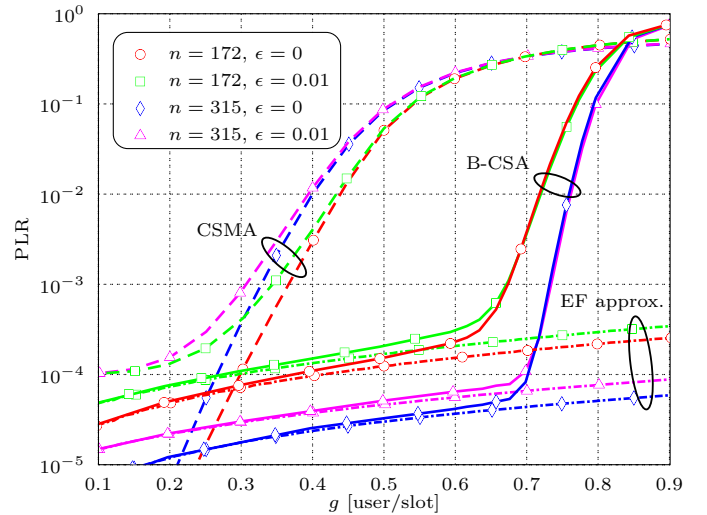}
	\caption{PLR comparison of optimized CSMA and B-CSA systems for different values of $n$ and $\epsilon$. Solid and the dashed lines show the performance of B-CSA and CSMA, respectively. The dash-dotted lines show the analytical PLR approximation obtained using~\eqref{eq:final_aprx}.}
	\label{fig:comparison}
\end{figure}

The broadcast PLR of the two protocols is shown in~\figref{fig:comparison} for $n = 172$ and $n = 315$ (see~\tabref{tab:params}). Simulation results for \gls{ABCSA} and \gls{CSMA} are shown with solid and dashed lines, respectively. The dash-dotted lines show the broadcast \gls{PLR} obtained using the approximation~\eqref{eq:final_aprx}. The figure shows that the protocols react differently to the increase of the frame length: The \gls{CSMA} performance degrades when $n$ increases whereas the performance of \gls{ABCSA} improves as $n$ grows large. This gives an extra degree of freedom when designing a \gls{ABCSA} system, as increasing the bandwidth will decrease the packet length and, hence, increase the number of slots. Moreover, \gls{ABCSA} is robust to packet erasures for any channel load as opposed to CSMA, which suffers significantly at low channel loads. We also point out that standard CSMA with $\kappa = 1$ fails in providing the required level of reliability over the PEC with $\epsilon = 0.01$ (see \figref{fig:csma_pec}).

It can also be seen from~\figref{fig:comparison} that \gls{ABCSA} significantly outperforms \gls{CSMA} for medium to high channel loads. For example, for $\epsilon = 0$, \gls{ABCSA} achieves a PLR of $10^{-3}$ at channel loads $g = 0.68$ and $g = 0.73$ for $n = 172$ and $n = 315$, respectively. \gls{CSMA} achieves the same reliability at $g = 0.36$ and $g = 0.32$ for $n = 172$ and $n = 315$, respectively, i.e., \gls{ABCSA} can support approximately twice as many users as \gls{CSMA} for this reliability. When erasures are present, the gains are even larger. For $\epsilon = 0$, CSMA yields better PLR than \gls{ABCSA} for low channel loads only ($g \lessapprox 0.3$). Moreover, \gls{CSMA} shows better performance for heavily loaded networks ($g>0.84$). However, in this case both protocols provide a poor reliability (\gls{PLR} of around $0.4$), which is unacceptable in~\glspl{VC}.

\revc{
We remark that the high user mobility prohibits the use of acknowledgements in \glspl{VC}, and thereby methods for mitigating the hidden terminal problem. Thus, the performance of \gls{CSMA} will be severely affected by the hidden terminal problem in a real vehicular network. On the other hand, the problem of hidden terminals does not exist in B-CSA since collisions are used for decoding and no sensing is required. 
}

\section{Conclusion and Discussion}
\revc{In this paper, we proposed a novel uncoordinated MAC protocol for a message exchange in the all-to-all broadcast scenario.} Furthermore, we analyzed its performance over the PEC for finite frame length and proposed an accurate analytical approximation of the PLR performance in the \gls{EF} region. The proposed analytical approximation can be used to optimize the degree distribution for CSA in the finite frame length regime. The analysis shows that \gls{ABCSA} is robust to packet erasures and is able to support a much higher number of users that can communicate reliably than the state-of-the-art MAC protocol currently used for \glspl{VC}.

In order to guarantee high reliability over the PEC, the \gls{PHY} layer and the MAC protocol of a communication system should offer a certain level of time diversity. For protocols that do not exploit collisions, such as \gls{CSMA}, increasing time diversity leads to channel congestion at low channel loads. Hence, exploiting collisions is inevitable for systems that require high reliability for high channel loads over the PEC. \gls{ABCSA} is an elegant way of utilizing collisions. The obtained gains come at the expense of increased computation complexity that a system designer should be ready to pay if they wants to satisfy the stringent requirements of \glspl{VC}.


\begin{thebibliography}{10}
\providecommand{\url}[1]{#1}
\csname url@samestyle\endcsname
\providecommand{\newblock}{\relax}
\providecommand{\bibinfo}[2]{#2}
\providecommand{\BIBentrySTDinterwordspacing}{\spaceskip=0pt\relax}
\providecommand{\BIBentryALTinterwordstretchfactor}{4}
\providecommand{\BIBentryALTinterwordspacing}{\spaceskip=\fontdimen2\font plus
\BIBentryALTinterwordstretchfactor\fontdimen3\font minus
  \fontdimen4\font\relax}
\providecommand{\BIBforeignlanguage}[2]{{%
\expandafter\ifx\csname l@#1\endcsname\relax
\typeout{** WARNING: IEEEtran.bst: No hyphenation pattern has been}%
\typeout{** loaded for the language `#1'. Using the pattern for}%
\typeout{** the default language instead.}%
\else
\language=\csname l@#1\endcsname
\fi
#2}}
\providecommand{\BIBdecl}{\relax}
\BIBdecl

\bibitem{Abramson70}
N.~Abramson, ``Another alternative for computer communications,'' in
  \emph{Proc. Fall Joint Computer Conf.}, 1970, pp. 281--285.

\bibitem{Roberts75}
L.~G. Roberts, ``{ALOHA} packet system with and without slot and capture,''
  \emph{SIGCOMM Comput. Commun. Rev.}, vol.~5, no.~2, pp. 28--42, Apr. 1975.

\bibitem{Choudhury83}
G.~Choudhury and S.~Rappaport, ``Diversity {ALOHA-A} random access scheme for
  satellite communications,'' \emph{{IEEE} Trans. Commun.}, vol.~31, no.~3, pp.
  450--457, Mar. 1983.

\bibitem{Casini07}
E.~Casini, R.~D. Gaudenzi, and O.~Herrero, ``Contention resolution diversity
  slotted {ALOHA} ({CRDSA}): An enhanced random access schemefor satellite
  access packet networks,'' \emph{{IEEE} Trans. Wireless Commun.}, vol.~6,
  no.~6, pp. 1408--1419, Apr. 2007.

\bibitem{Gollakota08}
S.~Gollakota and D.~Katabi, ``Zigzag decoding: Combating hidden terminals in
  wireless networks,'' \emph{{ACM} {SIGCOMM}}, pp. 159--170, 2008.

\bibitem{Tehrani11}
A.~Tehrani, A.~Dimakis, and M.~Neely, ``Sigsag: Iterative detection through
  soft message-passing,'' \emph{{IEEE} J. Sel. Top. Sign. Proces.}, vol.~5,
  no.~8, pp. 1512--1523, Dec. 2011.

\bibitem{ParandehGheibi10}
A.~ParandehGheibi, J.~K. Sundararajan, and M.~Medard, ``Collision helps -
  algebraic collision recovery for wireless erasure networks,'' in \emph{Proc.
  {IEEE} Wirel. Network Coding Conf.}, Boston, MA, June 2010.

\bibitem{Liva11}
G.~Liva, ``Graph-based analysis and optimization of contention resolution
  diversity slotted {ALOHA},'' \emph{IEEE Trans. Commun.}, vol.~59, no.~2, pp.
  477--487, Feb. 2011.

\bibitem{Narayanan12}
K.~R. Narayanan and H.~D. Pfister, ``Iterative collision resolution for slotted
  {ALOHA}: an optimal uncoordinated transmission policy,'' in \emph{Proc. Int.
  Symp. on Turbo Codes \& Iterative Information Processing}, Gothenburg,
  Sweden, Aug. 2012.

\bibitem{Paolini16}
E.~Paolini, G.~Liva, and M.~Chiani, ``Coded slotted {ALOHA}: A graph-based
  method for uncoordinated multiple access,'' \emph{{IEEE} Trans. Inf. Theory
  (to appear)}, 2016.

\bibitem{Stefanovic13}
C.~Stefanovic and P.~Popovski, ``{ALOHA} random access that operates as a
  rateless code,'' \emph{IEEE Trans. Commun.}, vol.~61, no.~11, pp. 4653--4662,
  Nov. 2013.

\bibitem{Olfati-Saber07}
R.~Olfati-Saber, J.~A. Fax, and R.~M. Murray, ``Consensus and cooperation in
  networked multi-agent systems,'' \emph{Proc. of the {IEEE}}, vol.~95, no.~1,
  pp. 215--233, Jan. 2007.

\bibitem{Popovski14}
P.~Popovski, ``Ultra-reliable communication in 5{G} wireless systems,'' in
  \emph{Proc. {IEEE} Int. Conf. 5{G} for Ubiquitous Connectivity}, Levi,
  Finland, Nov. 2014.

\bibitem{Johansson15}
N.~A. Johansson, Y.~Wang, E.~Eriksson, and M.~Hessler, ``Radio access for
  ultra-reliable and low-latency {5G} communications,'' in \emph{Proc. {IEEE}
  Int. Conf. Commun.}, London, UK, June 2015.

\bibitem{Durisi16}
G.~Durisi, T.~Koch, and P.~Popovski, ``Towards massive, ultra-reliable, and
  low-latency wireless communication with short packets,'' \emph{Proc. of the
  {IEEE} (to appear)}, 2016.

\bibitem{Di02}
C.~Di, D.~Proietti, I.~E. Telatar, T.~J. Richardson, and R.~L. Urbanke,
  ``Finite-length analysis of low-density parity-check codes on the binary
  erasure channel,'' \emph{IEEE Trans. Inf. Theory}, vol.~48, no.~6, pp.
  1459--1473, June 2002.

\bibitem{Wang09}
C.-C. Wang, S.~R. Kulkarni, and H.~V. Poor, ``Finding all small error-prone
  substructures in {LDPC} codes,'' \emph{{IEEE} Trans. Inf. Theor.}, vol.~55,
  no.~5, pp. 1976--1999, May 2009.

\bibitem{Rosnes09}
E.~Rosnes and O.~Ytrehus, ``An efficient algorithm to find all small-size
  stopping sets of low-density parity-check matrices,'' \emph{{IEEE} Trans.
  Inf. Theor.}, vol.~55, no.~9, pp. 4167--4178, Sep. 2009.

\bibitem{Orlitsky05}
A.~Orlitsky, K.~Viswanathan, and J.~Zhang, ``Stopping set distribution of
  {LDPC} code ensembles,'' \emph{IEEE Trans. Inf. Theory}, vol.~51, no.~3, pp.
  929--953, Mar. 2005.

\bibitem{Pao15}
E.~Paolini, ``Finite length analysis of irregular repetition slotted {ALOHA}
  ({IRSA}) access protocols,'' in \emph{Proc. IEEE Int. Conf. Commun.
  Workshop}, London, UK, June 2015.

\bibitem{Ivanov14}
M.~Ivanov, F.~Br\"{a}nnstr\"{o}m, A.~{Graell i Amat}, and P.~Popovski, ``Error
  floor analysis of coded slotted {ALOHA} over packet erasure channels,''
  \emph{IEEE Commun. Lett.}, vol.~19, no.~3, pp. 419--422, Mar. 2015.

\bibitem{Fabregas06}
A.~{Guill\'en i F\`abregas}, ``Coding in the block-erasure channel,''
  \emph{IEEE Trans. Inf. Theory}, vol.~52, no.~11, pp. 5116--5121, Nov. 2006.

\bibitem{Munari13}
A.~Munari, M.~Heindlmaier, G.~Liva, and M.~Berioli, ``The throughput of slotted
  {ALOHA} with diversity,'' in \emph{Proc. Allerton Conf. Commun., Control, and
  Computing}, Monticello, IL, Oct. 2013.

\bibitem{Herrero14}
O.~D.~R. Herrero and R.~D. Gaudenzi, ``Generalized analytical framework for the
  performance assessment of slotted random access protocols,'' \emph{IEEE
  Trans. Wireless Commun.}, vol.~13, no.~2, pp. 809--821, Feb. 2014.

\bibitem{Lottici02}
V.~Lottici, A.~D'Andrea, and U.~Mengali, ``Channel estimation for
  ultra-wideband communications,'' \emph{{IEEE} J. Select. Areas Commun.},
  vol.~20, no.~9, pp. 1638--1645, Dec. 2002.

\bibitem{Rimoldi96}
B.~Rimoldi and R.~Urbanke, ``A rate-splitting approach to the {Gaussian}
  multiple-access channel,'' \emph{{IEEE} Trans. Inf. Theory}, vol.~42, no.~2,
  pp. 364--375, Mar. 1996.

\bibitem{Luby98}
M.~G. Luby, M.~Mitzenmacher, M.~A. Shokrollahi, and D.~A. Spielman, ``Analysis
  of low density codes and improved designs using irregular graphs,'' in
  \emph{Proc. {ACM} Symp. on Theory of Computing}, 1998.

\bibitem{Richardson01}
T.~J. Richardson and R.~L. Urbanke, ``The capacity of low-density parity-check
  codes under message-passing decoding,'' \emph{IEEE Trans. Inf. Theory},
  vol.~47, no.~2, pp. 599--618, Feb. 2001.

\bibitem{Bondy76_Book}
J.~A. Bondy and U.~S.~R. Murty, \emph{Graph Theory with Applications},
  1st~ed.\hskip 1em plus 0.5em minus 0.4em\relax North-Holland, 1976.

\bibitem{Ivanov15}
M.~Ivanov, F.~Br\"{a}nnstr\"{o}m, A.~{Graell i Amat}, and P.~Popovski,
  ``All-to-all broadcast for vehicular networks based on coded slotted
  {ALOHA},'' in \emph{Proc. IEEE Int. Conf. Commun. Workshop}, London, UK, June
  2015.

\bibitem{Lubi98}
M.~Luby, M.~Mitzenmacher, and A.~Shokrollahi, ``Analysis of random processes
  via and-or tree evaluation,'' in \emph{Proc. {ACM-SIAM} Symp. Discrete
  Algorithms}, San Francisco, CA, Jan. 1998.

\bibitem{Skiena08_Book}
S.~S. Skiena, \emph{The Algorithm Design Manual}, 2nd~ed.\hskip 1em plus 0.5em
  minus 0.4em\relax Springer, 2008.

\bibitem{Lee05}
S.-R. Lee, S.-D. Joo, and C.-W. Lee, ``An enhanced dynamic framed slotted
  {ALOHA} algorithm for {RFID} tag identification,'' in \emph{Proc. Int. Conf,
  on Mobile and Ubiquitous Systems: Networking and Services}, San Diego, CA,
  July 2005.

\bibitem{Ryan_Lin_Book}
W.~E. Ryan and S.~Lin, \emph{Channel codes: Classical and Modern},
  1st~ed.\hskip 1em plus 0.5em minus 0.4em\relax Cambridge University Press,
  2009.

\bibitem{Boyd09_Book}
S.~Boyd and L.~Vandenberghe, \emph{Convex optimization}, 1st~ed.\hskip 1em plus
  0.5em minus 0.4em\relax Cambridge University Press, 2009.

\bibitem{Lagarias98}
J.~Lagarias, J.~A. Reeds, M.~H. Wright, and P.~E. Wright, ``Convergence
  properties of the {Nelder-Mead} simplex method in low dimensions,''
  \emph{SIAM Journal of Optimization}, vol.~9, no.~1, pp. 112--147, 1998.

\bibitem{IEEE80211}
{IEEE Std. 802.11-2012}, ``Part 11: Wireless {LAN} medium access control
  ({MAC}) and physical layer ({PHY}) specifications,'' Tech. Rep., Mar. 2012.

\bibitem{etsi_cam}
{ETSI EN 302 637-2: Draft V.0.0.5}, ``Intelligent transport systems ({ITS});
  {Vehicular} communications; {Basic} set of applications; {Part} 2:
  Specification of cooperative awareness basic service,'' Tech. Rep., June
  2012.

\bibitem{Bilstrup09}
K.~Bilstrup, E.~Uhlemann, E.~Str\"{o}m, and U.~Bilstrup, ``On the ability of
  the 802.11p {MAC} method and {STDMA} to support real-time vehicle-to-vehicle
  communication,'' \emph{EURASIP J. on Wireless Commun. and Networking}, vol.
  2009, pp. 1--13, Apr. 2009.

\end{thebibliography}
\end{document}